\documentclass[preprint]{elsarticle}

\journal{Journal of \LaTeX\ Templates}

\biboptions{authoryear}

\usepackage[title]{appendix}
\usepackage{amsmath}
\newtheorem{theorem}{Theorem}

\usepackage{graphicx}
\usepackage{times}
\usepackage[english]{babel}
\usepackage{epstopdf}
\usepackage{algorithm}
\usepackage{algorithmic}
\usepackage{booktabs}
\usepackage[font=small,labelfont=bf]{caption}
\usepackage{amsfonts}
\usepackage{subfigure}
\usepackage{natbib}
\DeclareMathOperator*{\argmin}{arg\,min}
\DeclareMathOperator*{\argmax}{arg\,max}
\usepackage{amssymb}

\begin{document}

\begin{frontmatter}

\title{Generalized Elliptical Slice Sampling with Regional Pseudo-priors}

\author[rvt]{Song LI\corref{cor1}}
\cortext[cor1]{Corresponding author}
\ead{sli228-c@my.cityu.edu.hk}

\author[rvt]{Geoffrey K. F. Tso}
\ead{msgtso@cityu.edu.hk}



\begin{abstract}
In this paper, we propose a MCMC algorithm based on elliptical slice sampling with the purpose to improve sampling efficiency. During sampling, a mixture distribution is fitted periodically to previous samples. The components of the mixture distribution are called regional pseudo-priors because each component serves as the pseudo-prior for a subregion of the sampling space. Expectation maximization algorithm, variational inference algorithm and stochastic approximation algorithm are used to estimate the parameters. Meanwhile, parallel computing is used to relieve the burden of computation. Ergodicity of the proposed algorithm is proven mathematically. Experimental results on one synthetic and two real-world dataset show that the proposed algorithm has the following advantages: with the same starting points, the proposed algorithm can find more distant modes; the proposed algorithm has lower rejection rates; when doing Bayesian inference for uni-modal posterior distributions, the proposed algorithm can give more accurate estimations; when doing Bayesian inference for multi-modal posterior distributions, the proposed algorithm can find different modes well, and the estimated means of the mixture distribution can provide additional information for the location of modes.
\end{abstract}


\end{frontmatter}


\section{Introduction}
The Markov Chain Monte Carlo (MCMC) algorithm was first proposed in the late 1940s, and has become popular in the recent decades because it can generate samples from any arbitrary complex target distribution. Denote the probability of state $i$ as $\pi_{i}$,  and the transition probability from state $i$ to state $j$ as $T_{ij}$. $\pi$ is called invariant or stationary with respect to (w.r.t) the Markov chain if the transition function $T$ leaves the distribution unchanged. For MCMC algorithm, this can be guaranteed by the detailed balance condition (DBC) $\pi_{i}p_{ij} = \pi_{j}p_{ji}$. The intuition of DBC is that the amount of probability mass leaving from state $i$ to state $j$ is the same as the amount leaving from state $j$ to state $i$. Two classic algorithms that satisfy DBC are the Metropolis-Hastings (MH)  (\cite{hastings1970monte,chib1995understanding}) and the Gibbs sampling algorithm (\cite{geman1984stochastic}).

\begin{itemize}
\item MH algorithm: denote the target distribution as $\pi(\mathbf{x})$, the starting point as $\mathbf{x_{0}} \in S$ and the proposal distribution as $q(\mathbf{y|x})$. At iteration $n$, a candidate point $\mathbf{y}$ is drawn from distribution $q(\mathbf{y}|\mathbf{x_{n-1}})$. This newly proposed candidate $\mathbf{y}$ is `accepted' as the next sample ($\mathbf{x_{n+1}}=\mathbf{y}$) with probability $\min\{ {1, \frac{\pi(\mathbf{y})q(\mathbf{x_{n}}|\mathbf{y})}{\pi(\mathbf{x_{n}})q(\mathbf{y}|\mathbf{x_{n}})}}\}$, or `rejected' ($\mathbf{x_{n+1}}=\mathbf{x_{n}}$) with probability $1-\min\{ {1, \frac{\pi(\mathbf{y})q(\mathbf{x_{n}}|\mathbf{y})}{\pi(\mathbf{x_{n}})q(\mathbf{y}|\mathbf{x_{n}})}}\}$. 
\item Gibbs sampling: Gibbs sampling is a special case of Metropolis-Hastings sampling. Suppose the target distribution is a multivariate joint distribution $\pi(x_{1},...,x_{n})$, Gibbs sampling samples each variable in turn from $p(x_{i}|x_{-i})$ (the distribution of that variable conditional on all the other variables). 
\end{itemize}

MCMC algorithm is easy to implement in the sense that it does not require too many mathematical derivations. However, the `easiness' is at the sacrifice of computational burden. Some of the inherent drawbacks of MCMC algorithm are the heavy computational burden, the poor performance when finding distant modes and difficulty dealing with strong dependency between variables. These drawbacks have greatly limited its popularization, especially in this era of big data. In recent years, many researchers have proposed many novel algorithms to relieve above problems. They are in the following categories. 1. algorithms evolved from physical dynamics, such as Hamiltonian MC (\cite{neal2011mcmc}), Langevin dynamics MC (\cite{welling2011bayesian}) and the bouncy particle sampler (\cite{bouchard2018bouncy}); 2. MCMC algorithms with adaptive transition kernels, such like \cite{atchade2006adaptive,wang2013adaptive}). 3. MCMC algorithms that take advantage of the development of hardwares (multi-core CPU and GPU), such like \cite{craiu2009learn,li2017powered} and \cite{white2014gpu}. 


The algorithm proposed in this paper is based on elliptical slice sampling (ESS) (\citeauthor{murray2010elliptical}). ESS is an extension of slice sampling which was proposed by \citeauthor{neal2003slice} in 2003. Next, we briefly introduce slice sampling and elliptical slice sampling. 

\textbf{Slice sampling}: sampling from a density distribution $f(x)$ is equivalent to first sampling from the region $\{(x,y):0 \leq y \leq f(x)\}$, then projecting the samples to the space of $x$. \citeauthor{neal2003slice} (2003) found that the first step can be done by alternating uniformly sampling along the vertical direction with uniformly sampling from the horizontal ``slice", that follows a trajectory relative to its position on the vertical axis. Slice sampling can adapt the step size of each variable and adapt to the dependency between variables based on their local properties, so as to increase the acceptance rates. Many variations of slice sampling were subsequently developed (\cite{tibbits2011parallel,liechty2010multivariate,kalli2011slice}).

\citeauthor{murray2010elliptical} (2010) proposed \textbf{elliptical slice sampling (ESS)} to sample from Gaussian prior models i.e. $\pi(x)=\frac{1}{Z}\mathcal{N}(x;0;\Sigma)L(x)$, where $\mathcal{N}(x;0;\Sigma)$ is the Gaussian prior distribution and $L(x) = p(DATA|x)$ is the likelihood function. At each iteration, a contour is constructed by the current sample and a new auxiliary sample generated by the Gaussian prior. The next sample is found on the contour after repeated rotations of $x$ until some criterion is met. The idea of slice sampling is used to reduce the choice of possible angles during rotation so as to increase the sampling efficiency. In some way, ESS algorithm converts a multi-dimensional sampling problem to a one-dimensional problem by sampling a series of angles. ESS has many advantages, such as good at dealing with variables dependencies, easy to implement, having no free parameters and performing well in both low and high dimensional settings. 


As shown above, slice sampling and ESS are efficient in many settings. However, they can get in trouble when the target distribution has distant modes. When sampling from multi-modal distributions, the samples could be trapped in some mode. We propose a \textbf{parallel adaptive generalized elliptical slice sampling with regional pseudo-priors (RGESS)} to relieve above issues. The advantages of RGESS includes: 1. ESS can only sample from Gaussian prior distributions, RGESS can sample from arbitrary target distributions; 2. the transition kernel is adapted on the fly to increase the sampling efficiency; 3. parallel computing is used to relieve the burden of calculation.

To our knowledge, we are the first to take advantage of combining parallel computing, elliptical slice sampling, mixture distribution adaption and regional pseudo-priors. Theoretically proofs are given to show the validity of our algorithms. Experimental results show that the proposed samplers have the following advantages: 1. with the same starting points, our algorithms can reach more distant modes; 2. Our algorithm has lower rejection rates; 3. When doing Bayesian inference for uni-modal target distributions, our algorithm can give more accurate estimations; 4. When doing Bayesian inference for multi-modal target distribution, our algorithm can find the modes well, and the estimated means of the mixture distribution can indicate the location of modes.

The structure of this paper is as follows. Some necessary prerequisite knowledge about slice sampling and elliptical slice sampling is given in section 2, details of our proposed samplers are given in section 3, theoretical results are shown in section 4, and the experimental results are presented in section 5.

\section{Prerequisites}
\subsection{Slice Sampling}
Denote the target distribution as $\pi(\mathbf{x}), \mathbf{x} \in R^{n}$. Instead of sampling $\mathbf{x}$ directly, slice sampling samples uniformly from the region $\mathcal{S}=\{ (\mathbf{x},y): 0\leq y\leq \pi(\mathbf{x}) \}$, then projects the samples to the space of $\mathbf{x}$. Denote the initial point as $\mathbf{x_{0}}$, which is randomly sampled from the sampling space. The procedures for sampling subsequent samples are:
\begin{enumerate}
\item Draw a real value $y_{0}$ from $p(y|\mathbf{x_{0}})$, which is the uniform distribution on $[0,f(\mathbf{x_{0}})]$. 
\item Define the horizontal ``slice": $S=\{{\mathbf{x}:y_{1}<f(\mathbf{x})}\}$. 
\item Draw a new point $\mathbf{x_{1}}$ from $p(\mathbf{x}|y_{0})$, which is the uniform distribution on the horizontal slices $S$ defined in step 2.
\item Repeat step 1 to 3.
\end{enumerate}
Figure 1 is an illustration of slice sampling. Slice sampling updates $\mathbf{x}$ and $y$ alternately from $p(\mathbf{x}|y)$ and $p(y|\mathbf{x})$ to leave the distribution $p(\mathbf{x},y)$ invariant. $p(y|\mathbf{x})$ is just the uniform distribution on the horizontal slice, but $p(\mathbf{x}|y)$ is nontrivial because the $S$ can be discontinuous. \citeauthor{neal2003slice} (2003) offered two methods, the stepping out and shrinkage procedure and the doubling procedure. The stepping out procedure first randomly finds an interval $w$ around the current point $x$ and expand the interval of size $w$ until both ends are outside $S$, then repeatedly uniformly sample a new point on the interval until the point is in $S$, those points outside $S$ are used to shrink the expanded interval. For the doubling procedure, the only difference is that the interval $w$ is repeatedly doubled until both ends are outside $S$. Please refer to \cite{neal2003slice} for more details. 

\begin{figure}[t]
\centering 
\includegraphics[height=2in]{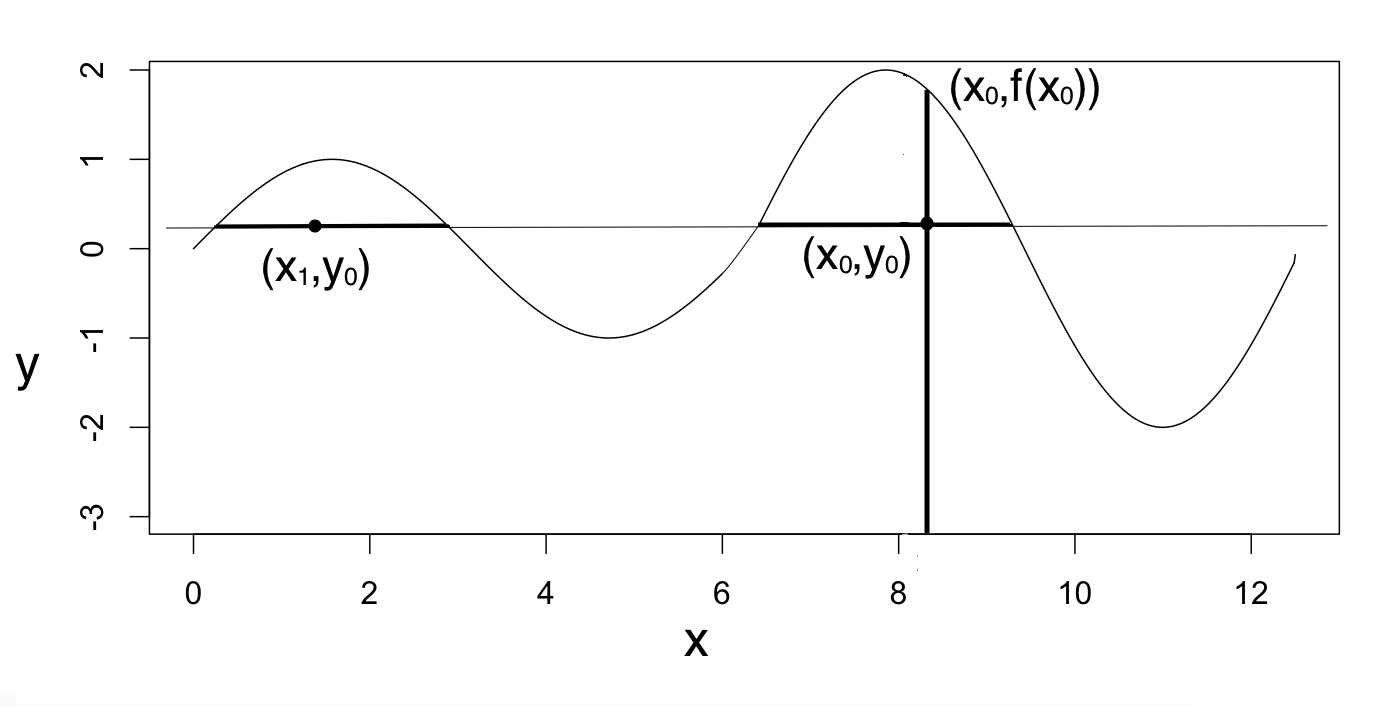}
\caption{Slice sampling: denote the target distribution as $\mathrm{f(x)}$, the initial point as $x_{0}$ is randomly sample from the sampling space. 1. Uniformly sample $\mathrm{y_{0}}$ from the vertical line $[0,f(\mathrm{x_{0}})]$; 2. uniformly sample $\mathrm{x_{1}}$ from the horizontal slice $S=\{\mathbf{x}:y_{1}<f(\mathbf{x})\}$; 3. start with $\mathrm{x_{1}}$ and repeat step 1 and 2. } \label{frontier}
\end{figure}

\subsection{Elliptical slice sampling}
\citeauthor{murray2010elliptical} (2010) proposed Elliptical Slice Sampling (ESS) as an extension of slice sampling to sample from models with Gaussian prior:
\begin{equation}
\pi(\mathbf{x}) \propto L(\mathbf{x})\mathcal{N}(\mathbf{x};0;\Sigma),
\end{equation}
where $\pi(\mathbf{x})$ is the posterior distribution, $L(\mathbf{x})$ is the likelihood function and $\mathcal{N}(\mathbf{x};0;\Sigma)$ is the multivariate Gaussian prior. Such models are also called \textit{latent Gaussian models} and have been used frequently in Gaussian processes and Gaussian Markov random fields. 

ESS algorithm is evolved from MH algorithm in \cite{bernardo1998regression} with the following proposal function: 
\begin{equation}
\mathbf{x}'=\sqrt{1-\epsilon^{2}}\mathbf{x}+\epsilon \mathbf{\nu}, ~~\mathbf{\nu} \sim \mathcal{N}(0,\Sigma)
\end{equation}
where $\mathbf{x}$ is the current point and $\mathbf{x}'$ is the candidate point, $\mathbf{\nu}$ is an auxiliary variable sampled from the Gaussian prior distribution. $\mathbf{x}'$ is accepted as the next state with probability $ \min\{ 1,  L(\mathbf{x}')/ L(\mathbf{x}) \} $ or the next state is a copy of $\mathbf{x}$. Above algorithm satisfies detailed balance with respect to Gaussian prior for $\mathbf{x}$. If we reparametrize $\epsilon=\sin \theta$, then above equation becomes
\begin{equation}
\mathbf{x'}=\mathbf{x}\cos \theta+\mathbf{\nu}\sin \theta.
\end{equation}
$\mathbf{x'}$ can be deemed as a rotation of $\mathbf{x}$ on the contour constructed by $\mathbf{x}$ and $\mathbf{\nu}$ with angle $\theta$. Here $\theta$ is a representation of the step size $\epsilon$. Transition kernel Equations (2) and (3) are actually equivalent to $x'=\mathcal{N}(x, \Sigma)$.

Neal stated that the selection of step-size $\epsilon$ in equation (2) or the $\theta$ in equation (3) is crucial for constructing an efficient Markov chain. ESS combines equation (3) with slice sampling to adaptively tune the step size. Given the current state $\mathbf{x}$, ESS first samples an auxiliary variable $v$ from the Gaussian prior to define the contour, and samples an angle $\theta$ uniformly from $0$ to $2\pi$. Similar to the slice sampling, a threshold is defined by $\log(y) = \log L(\mathbf{x}) + \log u$, where $u \sim Uniform(0,1)$. Next we want to find a state on the contour $C = \{\mathbf{x}' | \mathbf{x}' = (\mathbf{x}-\mathbf{\mu})cos(\theta)+(\mathbf{\nu}-\mathbf{\mu})sin\theta+\mathbf{\mu}, \theta \in [0, 2\pi] \}$ that also lies on the `slice' $\mathcal{S}=\{{\mathbf{x} : \log(y)<L(\mathbf{x})}\}$. The bracket of $\theta$ in $C$ is repeatedly shrinked when a new proposal is not on $C \cap S$ until the proposed $\mathbf{x}'$ is accepted. Above strategy combines the adaptive feature of slice sampling and the concept of `contour', thus is called elliptical slice sampling. Please refer to \cite{murray2010elliptical} for more details,. 

Essentially, ESS converts a multivariate sampling problem to a univariate sampling problem, making the sampling process easier to implement. More importantly, the Gaussian prior makes ESS capable of capture dependencies among variables. Details of the algorithm are given in Algorithm 1. One can see step 6 is just the MH fashion of rejection or acception, while the difference is that ESS shrinks the proposal space when a new proposal is rejected. \textbf{Essentially,, ESS is a special case of MH algorithm with adaptive step size.} \textbf{Figure 2} is an illustration of ESS. Given the current state $x_{i}$, an auxiliary point $v_{i}$ is drawn from the Gaussian prior, then $x_{i+1}$ is drawn on the contour constructed by $x_{i}$ and $v_{i}$. 

\begin{figure}[t]
\centering 
\includegraphics[height=2in]{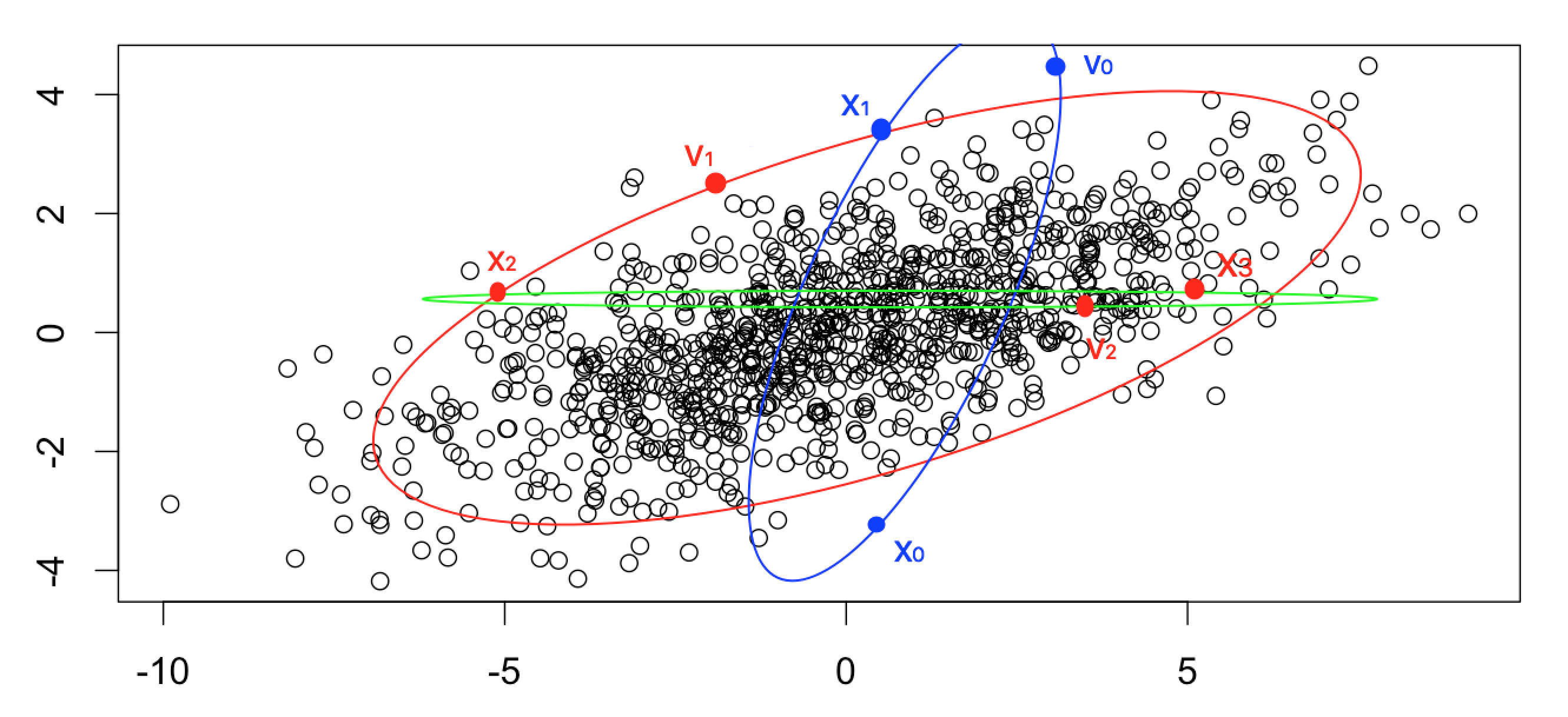}
\caption{Elliptical slice sampling. The target distribution is a two-dimensional Gaussian distribution centered at $(0, 0)$ with covariance matrix $[10,3; 3,2]$. For $i \in \mathbb{N}^{+}$: $\mathrm{v_{i}}$ are sampled from the Gaussian prior as in equation (2), which is the same as the target distribution; $\mathrm{x_{i+1}}$ are sampled from the contour constructed by $\mathrm{x_{i}}$ and $\mathrm{v_{i}}$. } \label{frontier}
\end{figure}

\begin{algorithm}
\caption{Elliptical slice sampling}
\textbf{Input}: current state $\mathbf{x}$, Gaussian prior parameters $\mathbf{\mu}$, $\Sigma$; log-likelihood function $log(L(\mathbf{x}))$.\\
\textbf{Output}: a new state $\mathbf{x'}$
\begin{algorithmic}[1]
\STATE Sample an auxiliary variable $\mathbf{v}$ to define the eclipse: $\mathbf{v} \sim \mathcal{N}(\mathbf{\mu},\Sigma)$
\STATE Find log-likelihood threshold: \\
$u \sim Uniform[0,1], log(y) \leftarrow logL(\mathbf{x}) + \log u$
\STATE $\theta \sim$ Uniform$[0,2\pi]$
\STATE $[\theta_{min}, \theta_{max}] \leftarrow [\theta-2\pi, \theta]$
\STATE $\mathbf{x'} \leftarrow (\mathbf{x}-\mathbf{\mu})cos(\theta)+(\mathbf{v}-\mathbf{\mu})sin\theta+\mathbf{\mu}$
\IF{$logL(\mathbf{x'}) > log(y)$}
\STATE return x'
\ELSE
\IF{$\theta<0$}
\STATE $\theta_{min} \leftarrow \theta$
\ELSE 
\STATE $\theta_{max} \leftarrow \theta$
\ENDIF
\STATE $\theta \sim Uniform[\theta_{min}, \theta_{max}] $
\STATE go to 5
\ENDIF
\end{algorithmic}
\end{algorithm}

\section{Adaptive GESS with regional pseudo-priors }
One limitation of elliptical slice sampling is that it can only be applied to Gaussian prior models. To generallize elliptical sampling to arbitrary distribution, \cite{fagan2016elliptical} decomposed the target density as the multiplication of a Gaussian distribution (pseudo prior) $\mathcal{N}$ and a residual function $R(x) = \frac{\pi}{\mathcal{N}}$. Similarly, \cite{nishihara2014parallel} decomposed an arbitrary target distribution $\pi(x)$ as the multiplication of a Student's t distribution (pseudo prior) $\mathcal{T}$ and a residual function $R(x) = \frac{\pi}{\mathcal{T}}$. Please refer to the two papers for more details.

When the target distribution is multi-modal, an uni-modal pseudo-prior can not approximate the target distribution well thus leading to sampling inefficiency. One good solution is to use different pseudo-priors at different regions. These pseudo-priors are called ``regional pseudo-priors'' because each of them only applies to a subregion. More specifically, our approach is to fit a mixture distribution $\sum_{m=1}^{M}f_{m}(x;\phi_{m})$ to previous samples periodically. Then given the current state, the component of the mixture distribution that has the largest probability is selected as the pseudo-prior to sample the next state. In other words, the sampling space $S$ is split into subregions $\{S_{m}\}_{m=1}^{M}$ as follows:
\begin{equation}
S_{m} = \{x:\argmax_{j}{f_{j}(x; \phi_{j})=m}\}
\end{equation} 
Where $f_{j}(x; \phi_{j})$ is the $j^{th}$ components of the fitted mixture distribution. For the states in region $S_{m}$, $f_{m}(x, \phi_{m})$ is used as the pseudo-prior for sampling in the next iteration. 


To make the proposed algorithm easier to understand, we first show MH sampling with regional proposals, then link it with elliptical slice sampling, finally show how to use mixture distribution to conduct regional generalized elliptical slice sampling (RGESS). Parallel computing and parameters estimation methods are given at last.

\subsection{Regional proposal for MH sampler}
For regional proposal MH sampler, different proposals are used in different regions. Denote the sampling space as $S$. Suppose the sampling space $S$ is split to $\{ S_{i} \}_{i=1}^{M} (\cup_{i} S_{i} = S)$ according to Equation (4), then $f_{m}$ is used as the proposal distribution when the current state is in $S_{m}$. This is equivalent to using transition kernel 
\begin{equation}
f(x'|x)=\sum_{i=1}^{M}\mathcal{I}(x\in S_{i})f_{i}(x')
\end{equation}

With equation (5) as the transition probability, Theorem 1 gives the acceptance rate for the regional MH algorithm. 

\begin{theorem}
Denote the acceptance rate of transiting from $x_{1}$ to $x_{2}$ as $A(x_{1}, x_{2})$. With target density $\pi$, proposal distributions $\{f_{i}\}_{i=1}^{n}$, if $A(x_{1},x_{2})$ satisfies:
\begin{equation}
A(x_{1}, x_{2})=\left\{
\begin{aligned}
&   \frac{\pi(x_{2})}{\pi(x_{1})}, & if &~ x_{1},x_{2} \in S_{i} \\
&   \frac{\pi(x_{2})f_{j}(x_{1})}{\pi(x_{1})f_{i}(x_{2})}, & if& ~ x_{2} \in S_{j}, x_{1} \in S_{i} \\
\end{aligned}
\right.
\end{equation}
the detailed balance of regional proposal MH sampler is satisfied.
\end{theorem}

According to equation (2) and (3), elliptical slice sampling algorithm is essentially the same as MH algorithm with proposal function $y' = \sqrt{1 - \epsilon^{2}} y + \epsilon v$, where $v\sim \mathcal{N}(0,\Sigma)$. Notice that the transition kernel $y'\sim \mathcal{N}(y,\Sigma)$ is the same as distribution of $v$, thus one can change the distribution from which $v$ is sampling from to change the transition kernel. Note that in step 6 of Algorithm 1,  $\log L(x') > \log(y) = \log L(x) + \log u$ can also be expressed as $\log \frac{L(x')}{L(x)} > u$. The later expression is just the MH fashion to accept or reject the new proposal. Knowing above relationship between MH algorithm and ESS algorithm, we can extend the regional MH algorithm to regional generalized elliptical slice sampling algorithm. To make the Markov chain satisfy global balance with regional pseudo priors, the statement in step 2 of Algorithm 1 should be changed to $\log(y) \gets \log L(x) + \log f_{i}(y|x) - \log f_{j}(x|y)$ if $x \in S_{i}$ and $y \in S_{j}$. In this way, theorem 1 can be generalized to regional generalized elliptical slice sampling. 

\subsection{Gaussian mixture regional generalized elliptical slice sampling}
In this subsection, we fit a mixture distribution from previous samples periodically to improve sampling efficiency. \cite{fagan2016elliptical} also used Gaussian distribution as the pseudo-prior to conduct generalized elliptical slice sampling. Suppose the current state is in $S_{i}$, the Gaussian component $\mathcal{N}(x|\mu_{i},\Sigma_{i})$ is selected from the Gaussian mixture distribution as the pseudo-prior, the target distribution $\pi$ can be rewritten as:
\begin{equation}
\pi(x) = R(x) \mathcal{N}(x|\mu_{i},\Sigma_{i})
\end{equation}
where $R(x) = \frac{\pi(x)}{\mathcal{N}(x|\mu_{i},\Sigma_{i})}$ is the residual function and plays the same role as $\mathcal{L}(x)$ in equation (1).
It is important to notice that the pseudo-prior in GESS actually plays the same role as the transition probability in MH algorithm. Thus, the following theorem follows Theorem 1.

\begin{theorem}
Suppose the partition of $S = \bigcup_{i=1}^{M}S_{i=1}^{M}$ and the Gaussian mixture distribution parameters $\{ \mathbf{\mu}_{i}$, $\Sigma_{i} \}_{i=1}^{M}$ are given. Denote $R_{i}(x) = \frac{\pi(x)}{\mathcal{N}(x|\mu_{i},\Sigma_{i})}$ as the residual function w.r.t Gaussian pseudo-prior $\mathcal{N}(x|\mu_{i},\Sigma_{i})$. Pseudo-prior $ \mathcal{N} (x; \mu_{i}, \Sigma_{i})$ is used when $x \in S_{i}$.
With acceptance rate
\begin{equation}
A(x_{1}, x_{2})=\left\{
\begin{aligned}
&   \frac{\pi(x_{2})}{\pi(x_{1})}, & if &~ x_{1},x_{2} \in S_{i} \\
&   \frac{R_{i}(x_{2})}{R_{j}(x_{1})}, & if& ~ x_{2} \in S_{j}, x_{1} \in S_{i} \\
\end{aligned}
\right.
\end{equation}
the detailed balance of the GMRGESS is satisfied.
\end{theorem}
\centerline{(Proof is given in the appendix A2.)}
GMRGESS algorithm is given in Algorithm 2. Compared with elliptical slice sampling algorithm (Algorithm 1), the step 6 of Algorithm 2 is adjusted according to Theorem 2 to make GMRGESS satisfy detailed balance condition.

\begin{algorithm}
\caption{Gaussian mixture regional generalized elliptical slice sampling}
\textbf{Input}: current state $\mathbf{x}$ (suppose $\mathbf{x} \in S_{I}$); target function $\pi(\mathbf{x})$; Gaussian mixture distribution parameters $\{ \mathbf{\mu_{i}}$, $\Sigma_{i} \}_{i=1}^{M}$; $R_{i}(x) = \frac{\pi(x)}{\mathcal{N}(\mu_{i}, \Sigma_{i})}$. \\
\textbf{Output}: a new state $\mathbf{x'}$
\begin{algorithmic}[1]
\STATE Sample an auxiliary variable $\mathbf{v}$ to define the ellipse: $\mathbf{v} \sim \mathcal{N}(\mathbf{\mu}_{I},\Sigma_{I})$
\STATE $\theta \sim$ Uniform$[0,2\pi]$, $u \sim Uniform[0,1]$.
\STATE $[\theta_{min}, \theta_{max}] \leftarrow [\theta-2\pi, \theta]$
\STATE $\mathbf{x'} \leftarrow (\mathbf{x}-\mathbf{\mu_{I}})\cos(\theta)+(\mathbf{v}-\mathbf{\mu_{I}})\sin\theta+\mathbf{\mu}$, suppose $\mathbf{x'} \in S_{J}$ 
\STATE Find log-likelihood threshold: \\
$log(y) = \log(R_{J}(x)) + \log u$
\IF{$log R_{I}(\mathbf{x'}) > log(y)$}
\STATE return x'
\ELSE
\IF{$\theta<0$}
\STATE $\theta_{min} \leftarrow \theta$
\ELSE 
\STATE $\theta_{max} \leftarrow \theta$
\ENDIF
\STATE $\theta \sim Uniform[\theta_{min}, \theta_{max}] $
\STATE go to 4
\ENDIF
\end{algorithmic}
\end{algorithm}

\subsection{Student's t-mixture regional generalized elliptical sampling}
In this subsection, Student's t-mixture distribution is used because Student's t-distribution has longer tails than Gaussian distribution. \cite{nishihara2014parallel} used Student's t-distribution as the pseudo-prior to conduct generalized elliptical slice sampling. Inspired by their work, we fit a Student's t-mixture distribution to the previous samples periodically, select one component as the pseudo-prior to conduct Algorithm 2 in \cite{nishihara2014parallel}. We call this approach Student's t-mixture regional generalized elliptical sampling (TMRGESS). Similar to GMRGESS, the rejection rate should be adjusted to satisfy detailed balance condition which is shown in Theorem 3.


\begin{theorem}
Suppose the partition of $S = \bigcup_{i=1}^{M}S_{i=1}^{n}$ and the Student's t-mixture distribution parameters $\{ \mathbf{\mu}_{i}$, $\Sigma_{i}, \nu_{i} \}_{i=1}^{M}$ are given. Denote $R_{i}(x) = \frac{\pi(x)}{\mathcal{T}(x|\mu_{i},\Sigma_{i}, \nu_{i})}$ as the residual function. Pseudo-prior $ \mathcal{T} (x; \mu_{i}, \Sigma_{i}, \nu_{i})$ is used when $x \in S_{i}$. With acceptance rate
\begin{equation}
A(x_{1}, x_{2})=\left\{
\begin{aligned}
&   \frac{\pi(x_{2})}{\pi(x_{1})}, & if &~ x_{1},x_{2} \in S_{i} \\
&   \frac{R_{i}(x_{2})}{R_{j}(x_{1})}, & if& ~ x_{2} \in S_{j}, x_{1} \in S_{i} \\
\end{aligned}
\right.
\end{equation}
the detailed balance condition of the TMRGESS is satisfied.
\end{theorem}
\centerline{(Proof is given in the appendix A3.)}
The Student's t-mixture regional generalized elliptical sampling (TMRGESS) algorithm is given in Algorithm 3.

\begin{algorithm}
\caption{Student's t-mixture Regional generalized elliptical slice sampling}
\textbf{Input}: current state $\mathbf{x}$ (suppose $\mathbf{x} \in S_{I}$); dimension $D$; target function $\pi(\mathbf{x})$; Student's t-mixture distribution parameters $\{ \mathbf{\mu_{i}}$, $\Sigma_{i}, \nu_{i} \}_{i=1}^{M}$; $R_{i}(x) = \frac{\pi(x)}{\mathcal{T}(\mu_{i}, \Sigma_{i}, \nu_{i})}$ \\
\textbf{Output}: a new state $\mathbf{x'}$
\begin{algorithmic}[1]
\STATE $\alpha' \gets \frac{D+\nu_{I}}{2}$, $\beta' \gets \frac{1}{2}(\nu_{I} + (\mathbf{x}-\mu_{I})^{T}\Sigma_{I}^{-1}(\mathbf{x}-\mu_{I}))$
\STATE $s \sim IG(\alpha', \beta')$
\STATE Sample an auxiliary variable $v$ to define the ellipse: $\mathbf{v} \sim \mathcal{N}(\mathbf{\mu}_{I},s\Sigma_{I})$
\STATE $\theta \sim$ Uniform$[0,2\pi]$, $u \sim Uniform[0,1]$.
\STATE $[\theta_{min}, \theta_{max}] \leftarrow [\theta-2\pi, \theta]$
\STATE $\mathbf{x'} \leftarrow (\mathbf{x}-\mathbf{\mu_{I}})\cos(\theta)+(\mathbf{v}-\mathbf{\mu_{I}})\sin\theta+\mathbf{\mu}, ~\mathbf{x'} \in S_{J}$ 
\STATE Find log-likelihood threshold: \\
$log(y) = \log(R_{J}(x)) + \log u$
\IF{$log R_{I}(\mathbf{x'}) > log(y)$}
\STATE return x'
\ELSE
\IF{$\theta<0$}
\STATE $\theta_{min} \leftarrow \theta$
\ELSE 
\STATE $\theta_{max} \leftarrow \theta$
\ENDIF
\STATE $\theta \sim Uniform[\theta_{min}, \theta_{max}] $
\STATE go to 6
\ENDIF
\end{algorithmic}
\end{algorithm}

\subsection{Parallel computing and parameters estimation}
This section shows how to do the parameters adaption and parallel computing. To fit the Gaussian mixture distribution (section 3.2), we use three method: the expectation maximization (EM) algorithm, the variational inference (VI) algorithm and the stochastic approximation (SA) algorithm. To fit the Student's t-mixture distribution (section 3.3), we use expectation maximization (EM) algorithm only.

\subsubsection{Multi-chain Parallel}
One common problem of MCMC algorithm is that it usually requires heavy computation, therefore multiple chain parallel is ideal to relieve the burden of calculation. What's more, the use of multiple chains can also help discover the different modes of $\pi$. Suppose we run $K$ independent Markov chains in parallel. Denote $\mathcal{X}^{n}=\{X_{1}^{n},...,X_{K}^{n}\}$ as all the samples at iteration $n$, where $X_{k}^{n}$ is the sample at iteration $n$ on the $k^{th}$ chain. At iteration $n+1$, the parameters of Gaussian mixture distribution and Student's-t mixture distribution are estimated from $\mathcal{X}^{n}$ and used for the next iteration. An illustration is shown in Figure 3, where $\phi^{n}$ denotes the parameters estimated by samples in iteration $n$. The induced parallel version of adaptive regional generalized elliptical slice sampling is shown in \textbf{Algorithm 4}. In practice, we use StarCluster to launch clusters on Amazon EC2 to realize parallel computing.

\begin{figure}[h]
\centering 
\includegraphics[height=3in]{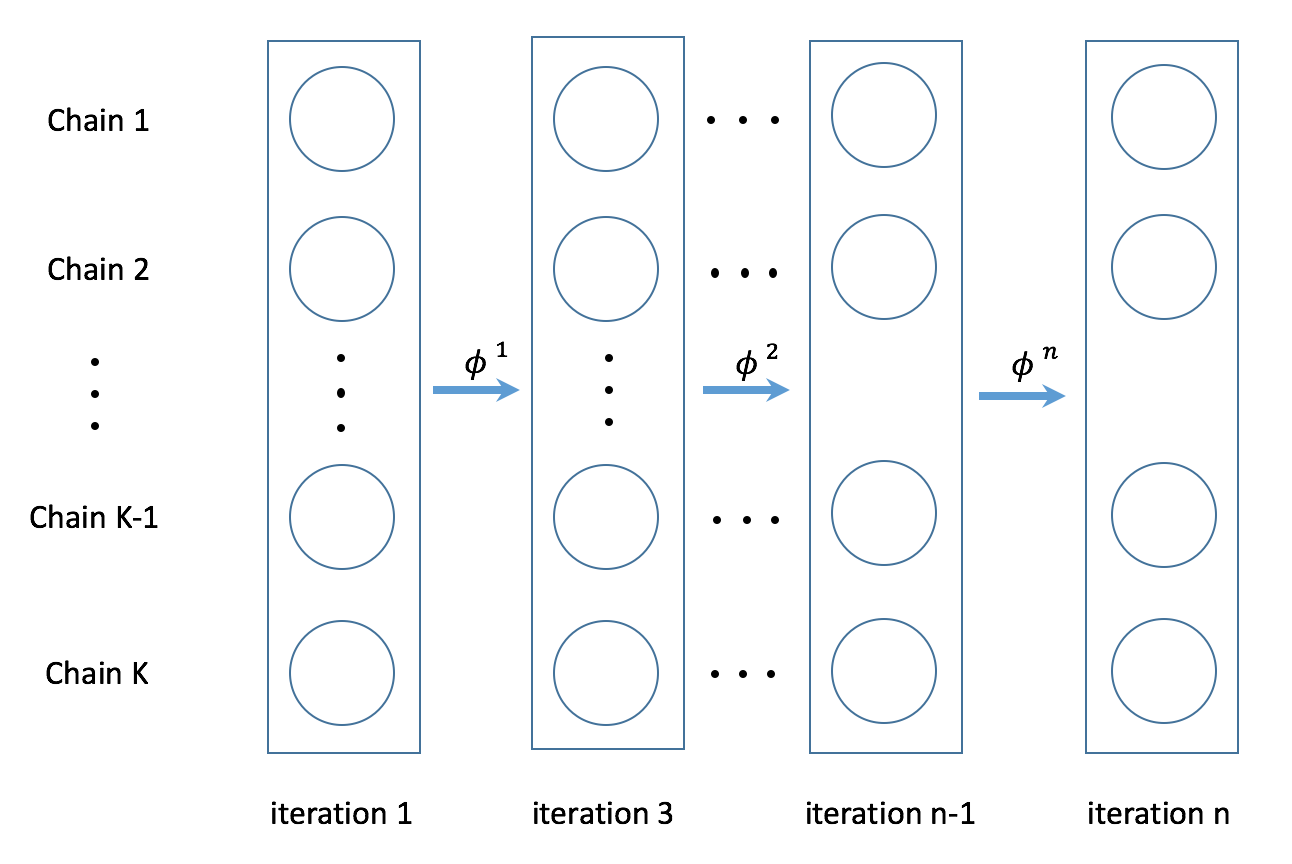}
\caption{Multi-chain parallel and parameters adaption. } \label{frontier}
\end{figure}

\begin{algorithm}
\textbf{Input}: $\phi^{0}=\{ {\mathbf{\mu^{0}},\Sigma^{0},w^{0}} \},  \mathcal{X}^{0}=\{X_{1}^{0},...,X_{K}^{0}\}$
\caption{Parallel Global Adaptive Generalized Elliptical Slice Sampling with Regional pseudo-priors}
\begin{algorithmic}[1]
\FOR{n=1,...,N}
\STATE Update the parameters $\phi^{n+1}=\{ {\mu^{n},\Sigma^{n},w^{n}} \}$ using adaption methods in section 3.4.2-3.4.4 to fit $\mathcal{X}^{n-1}$;
\STATE Next process is working on K threads simultaneously.
\FOR{$k=1,..., K$} 
\STATE At thread k, draw $X_{k}^{n}$ using  \textbf{Algorithm 2} or \textbf{Algorithm 3} from $X_{k}^{n-1}$. 
\ENDFOR
\STATE $\mathcal{X}^{n}=\{X_{1}^{n},...,X_{K}^{n}\}$. 
\ENDFOR
\end{algorithmic}
\end{algorithm}

\subsubsection{Expectation maximization algorithm}
Expectation maximization (EM) algorithm can iteratively approximate the maximum of likelihood function to estimate the parameters in statistical models with unobserved latent variables. Therefore, it is widely used to estimate the parameters of mixture distribution. Denote the unknown parameters at iteration $t$ as $\theta^{n}$, the observed data as $X$, the unobserved data as $Z$ and the likelihood function as $L(\theta;X,Z) = p(X,Z|\theta)$. EM algorithm includes two steps: the Expectation step (E step) to calculate the expected value of log likelihood with respect to the posterior distribution of $Z$ $ Q(\theta|\theta^{t}) = E_{Z|X,\theta^{t}}[\log L(\theta;X,Z)] $ and the Maximum step (M step) to maximize the expected log likelihood in the E step $\theta^{t+1} = \argmin_{\theta} Q(\theta|\theta^{t})$. Above two steps are iterated until the distance between $\theta^{t}$ and $\theta^{t+1}$ is smaller than a threshold. When applied to the mixture distribution, the unobserved data $Z$ is an indicator indicating which cluster $X$ belongs to. Details of applying EM algorithm to Gaussian mixture distribution can be found in \cite{bilmes1998gentle}, and details of applying EM algorithm to Student's t-mixture distribution can be found in  \cite{peel2000robust}. 

\subsubsection{Variational inference algorithm}
The EM algorithm in section 3.4.2 requires computing the expected value of log likelihood with respect to the posterior distribution of latent variable. However, in many practical cases it is infeasible or inefficient to conduct above calculations for many reasons, such as high dimensionality of latent space, large size of observations and high complexity of the posterior distribution. Therefore, an approximation approach is needed in these situations. Variational inference (VI) or variational Bayes algorithm is an approximate approach to minimize the KL divergence between a restricted family of distribution and the posterior distribution. Denote the observations as $X=\{ X_{1},...,X_{n} \}$, the unobserved data and unknown parameters as $Z=\{ Z_{1},...,Z_{n} \}$, the joint probability as $p(X,Z)$ and the posterior distribution as $p(Z|X)$. The log probability of $X$ can be decomposed as $\log p(X)=\mathcal{L}(q)+KL(q||p)$, where $\mathcal{L}(q)=\int q(Z) \log {\frac{p(X,Z)}{q(Z)}}dz$ is the lower bound, $KL(q||p)=-\int q(Z) \log{\frac{p(Z|X)}{q(Z)}}dZ$ is the KL divergence. Maximizing the lower bound $\mathcal{L}(q)$ is the same as minimizing the KL divergence between $q$ and $p$. Using mean field theory (\cite{parisi2010mean}) the form of $q(Z)$ is restricted to $q(Z)=\prod_{i}^{M}q_{i}(Z_{i})$. We optimize $\mathcal{L}(q)$ by optimizing with respect to each $q_{i}(Z_{i})$ in turn until some criteria is met. In this paper we follow the method in section 10.2 of \cite{bishop2006pattern} where variational inference is applied to Gaussian mixture distribution estimation.

\subsubsection{Stochastic approximation algorithm}
The EM algorithm and VI algorithm are frequently used to estimate parameters of mixture distributions. However, there are some practical issues when applying them to our algorithms. One problem is that the EM algorithm and VI algorithm are not robust to outliers. Figure 4 is a plot of points in two dimensional space fitted by Gaussian mixture distribution. There are two outliers on the right side. The red contours are the correct Gaussian mixture distribution, the blue contours are the results estimated by EM algorithm. The outliers are estimated as an independent Gaussian component whose covariance has small eigenvalues. Same problem also occurs to VI algorithm. When applied to generalized elliptical slice sampling, the biased estimation of Gaussian mixture distribution can further lead to biased samples in the next iteration. We would like to make the current estimation also influenced by previous estimations so as to reduce the effect of current outliers. In this way, inspired by the VI algorithm, we use stochastic approximation (SA) algorithm to gradually approach the correct values by minimizing the KL divergence between the Gaussian mixture distribution and the target distribution. Denote the samples at iteration $n$ as $X^{n}=\{ X^{n}_{1},...,X^{n}_{K} \}$, the estimated parameters of Gaussian mixture distribution at iteration $n$ as $ w^{n}=\{ w^{n}_{1},...,w^{n}_{M} \}, \mu^{n} = \{ \mu_{1}^{n},...,\mu_{M}^{n} \}, \Sigma^{n}=\{ \Sigma_{1}^{n},...,\Sigma_{M}^{n} \} $, the learning rate at iteration n as $r_{n}$. Using SA algorithm, the estimations are updated at iteration $n+1$ as follows:
\begin{equation}
\begin{split}
w_{j}^{n+1}=&w_{j}^{n}+r_{n+1}[\frac{1}{N}\sum_{k=1}^{K}\frac{\mathcal{N}(X^{n}_{k}|\mu_{j}^{n},\Sigma_{j}^{n})}{\sum_{m=1}^{M}w_{m}\mathcal{N}(X^{n}_{k}|\mathbf{\mu}_{m}^{n},\Sigma_{m}^{n})}\\
&-\frac{1}{MK}\sum_{k=1}^{K}\sum_{m=1}^{M}\frac{\mathcal{N}(X^{k}_{n}|\mu_{m}^{n},\Sigma_{m}^{n})}{\sum_{i=1}^{M}w_{i}^{n}\mathcal{N}(X^{n}_{k}|\mathbf{\mu}_{i}^{n},\Sigma_{i}^{n})}]
\end{split}
\end{equation}
\begin{equation}
\begin{split}
\mu_{j}^{n+1}=&\mathbf{\mu}_{j}^{n}+r_{n+1}\frac{1}{K}\sum_{k=1}^{K}\frac{\mathcal{N}(X^{n}_{k}|\mu_{j}^{n},\Sigma_{j}^{n})}{\sum_{m=1}^{M}w_{m}^{n}\mathcal{N}(X^{n}_{k}|\mathbf{\mu}_{m}^{n},\Sigma_{m}^{n})}\times \\& (\Sigma_{m}^{n})^{-1}(X^{n}_{k}-\mu_{m}^{n})
\end{split}
\end{equation}
\begin{equation}
\begin{split}
\Sigma_{j}^{n+1}=&\Sigma_{j}^{n}+r_{n+1}\frac{1}{K}\sum_{k=1}^{K}\frac{\mathcal{N}(X^{n}_{k}|\mu_{j}^{n},\Sigma_{j}^{n})}{\sum_{m=1}^{M}w_{m}^{n}\mathcal{N}(X^{n}_{k}|\mathbf{\mu}_{m}^{n},\Sigma_{m}^{n})}\times \\&[(X^{n}_{k}-\mu_{m}^{n})(X_{k}^{n}-\mu_{m}^{n})^{T}-\Sigma_{m}^{n}]
\end{split}
\end{equation}
Details can be found in Appendix B. Since the estimations are updated from previous estimations, they are less affected by outliers. 

\begin{figure}[h]
\centering 
\includegraphics[height=3in]{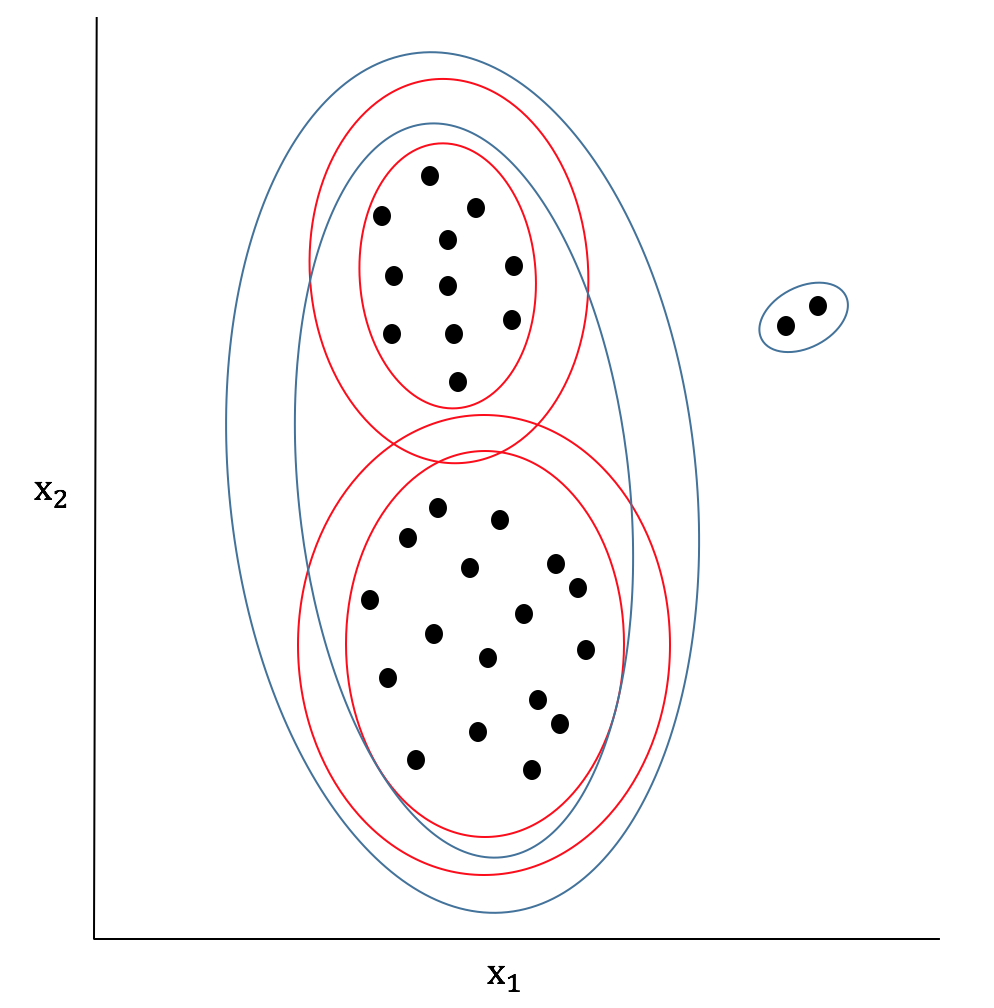}
\caption{Plot of points in two dimensional space fitted by Gaussian mixture distribution, there are two outliers on the right side. The red contours are the correct Gaussian mixture distribution, the blue contours are the results of EM algorithm. } \label{frontier}
\end{figure}


\subsubsection{Remarks}
As to be shown in section 4, as long as the estimated parameters satisfy the simultaneous uniform ergodicity and diminishing adaption conditions, the adaptive MCMC algorithm is valid. Therefore, we can use some tricks in practice that does not influence the validity of the algorithm. To facilitate exploring distant modes and avoid generating singular covariance matrix, in practice we can add a fixed diagonal matrix to the estimated covariance, a.e. $\tilde{\Sigma} + rI_{d}, r>0$.  The number of components is set manually even though determining the number of components of mixture models have been studied extensively (\cite{lo2001testing, mclachlan1987bootstrapping}). Since our purpose is to conduct efficient sampling, determining the number of components should not be a burden for our algorithm. Therefore, we simply use a relatively large number of components. In practice, we also found that sometimes the estimated covariance matrix is not positive semi-definite. We use the method in \cite{higham1988computing} to find the nearest positive semi-definite matrix and replace the original one. 

\section{Theoretical results}
Denote $P_{\gamma}^{N}$ as the transition kernel at iteration $N$ with adaption index $\gamma$. With fixed kernel $P_{\gamma}$ for $\gamma \in \mathcal{Y}$, it is shown in section 3 that the proposed algorithms satisfy detailed balance condition and are ergodic to $\pi(\cdot)$. Now we want to prove that the adaption preserves the ergodicity. Theorem 5 of \cite{roberts2007coupling} shows that an adaptive MH algorithm is ergodic as long as it satisfies the following two conditions: (a) [Simultaneous uniform ergodicity] For all $\epsilon > 0$, there is $N=N(\epsilon)\in \mathbf{N}$ such that $|| P_{\gamma}^{N}(x, \cdot) - \pi(\cdot) || \leq \epsilon$ for all $x\in \mathcal{X}$ and $\gamma \in \mathcal{Y}$; and (b) [Diminishing adaption] $\lim_{n \to \infty}\sup_{x \in \mathcal{X}} ||P_{\Gamma_{n+1}}(x,\cdot) - P_{\Gamma_{n}}(x,\cdot) || = 0$ in probability. We have stated in previous sections that the GESS algorithm is essentially the same as MH algorithm with the residual function as the proposal function. Therefore, ergodicity of the proposed algorithm follows the Theorem 5 of \cite{roberts2007coupling}. For writing convenience, next we introduce the abbreviations. For Gaussian mixture regional generalized elliptical slice sampling, \textbf{EM-GMRGESS} stands for using expectation maximization algorithm to estimate the Gaussian mixture distribution parameters, \textbf{VI-GMRGESS}  stands for using variational inference algorithm and \textbf{SA-GMRGESS} stands for using stochastic approximation algorithm. \textbf{EM-TMRGESS} stands for the Student's t-mixture regional generalized elliptical slice sampling using EM algorithm to estimate the parameters. The following theorems show the ergodicity of above algorithms.
\begin{theorem}
EM-GMRGESS satisfies simultaneous uniform ergodicity and diminishing adaption conditions, thus is ergodic to $\pi(\cdot)$.
\end{theorem}
\begin{theorem}
VI-GMRGESS satisfies simultaneous uniform ergodicity and diminishing adaption conditions, thus is ergodic to $\pi(\cdot)$.
\end{theorem}
\begin{theorem}
SA-GMRGESS satisfies simultaneous uniform ergodicity and diminishing adaption conditions, thus is ergodic to $\pi(\cdot)$.
\end{theorem}
\begin{theorem}
EM-TMRGESS satisfies simultaneous uniform ergodicity and diminishing adaption conditions, thus is ergodic to $\pi(\cdot)$.
\end{theorem}
\centerline{(Proofs are given in the appendix C.)}

\section{Experiments}
In this section, we first test the proposed algorithm by sampling from a Gaussian mixture distribution, then we apply them to a uni-modal and a multi-modal Bayesian inference problems using real-world datasets.

\subsection{Gaussian mixture model}
In this subsection, the algorithm is applied to a four-components Gaussian mixture target distribution as follows:
\begin{equation}
\pi(x) = w_{1}\mathcal{N}(x|\mu_{1},\Sigma_{1})+w_{2}\mathcal{N}(x|\mu_{2},\Sigma_{2})+w_{3}\mathcal{N}(x|\mu_{3},\Sigma_{3})+w_{4}\mathcal{N}(x|\mu_{4},\Sigma_{4})
\end{equation}
\begin{equation}
w_{1}=0.25, ~w_{2}=0.25,~w_{3}=0.25,~w_{4}=0.25
\end{equation}
\begin{equation}
\mu_{1}=[25,50],~\mu_{2}=[5,5], ~\mu_{3}=[50,5],~\mu_{4}=[50,50]
\end{equation}
\begin{equation}
\Sigma_{i}=10I_{2},
\end{equation}
 where $\mathnormal{I}_{4}$ denotes the $4 \times 4$ identity matrix.\\

Samples drawn from EM-GMRGESS, VI-GMRGESS and SA-GMRGESS algorithms are shown in Figure 5. The starting points are drawn from Gaussian distribution centered at $(20,20)$ with covariance $40I_{2}$, where $I_{2}$ denotes $2 \times 2$ identity matrix. In this example, we use 50 parallel chains and 4-components mixture. As shown in Figure 5, samples drawn from EM-GMRGESS and VI-GMRGESS algorithms are trapped in local modes because there are no connections among different modes. Samples drawn from SA-GMRGESS algorithm mix best, but with higher rejection rates. Here the rejection rates are defined as the number of rejection times before getting accepted. In practice, we find there is a trade-of between the rejection rates and the ability to explore distant modes. The reason is that Gaussian distributions have short tails. Covariance matrices with larger eigenvalues facilitate exploring distant modes, but in turn lead to higher rejection rates. This issue is also stressed in \cite{fagan2016elliptical}. One can notice that there are more outliers for EM-GMRGESS and VI-GMRGESS algorithms compared with SA-EMRGESS, which implies stochastic approximation algorithm can effectively reduce the influence of outliers.

\begin{figure}[h]
\centering 
\includegraphics[height=2.5in]{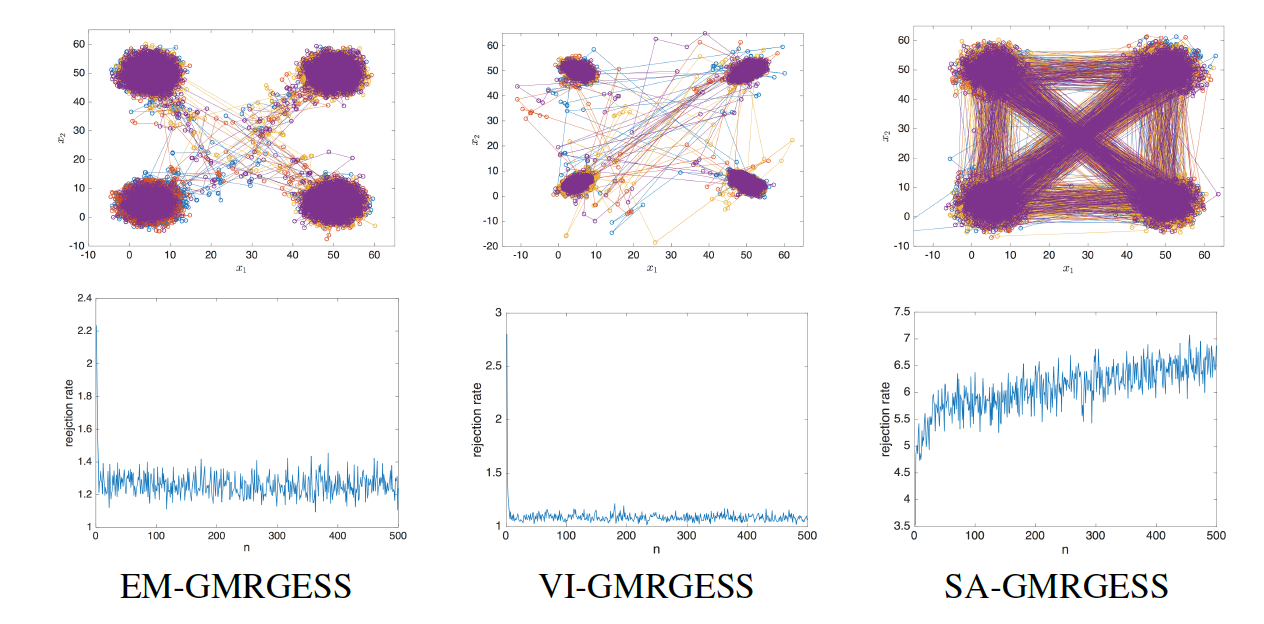}
\caption{From left to right: EM-GMRGESS algorithm, VI-GMRGESS algorithm and SA-GMRGESS algorithm. Plots of the first row show the drawn samples, in which two consecutive samples are connected with a line (0-125 iterations: blue; 125-250 iterations: brown; 250-375 iterations: orange; 375-500 iterations: purple). Plots of the second row show the rejection rates as a function of iteration numbers. } \label{frontier}
\end{figure}

Figure 6 shows the comparison of GESS algorithm in \cite{nishihara2014parallel} and the EM-TMRGESS algorithm, both with starting points drawn from $\mathcal{N}((5,5), 5 \times I_{2})$. For EM-TMRGESS, we still use 50 parallel chains and 4-component mixture. GESS totally misses the other modes except the mode around the starting points, while EM-TMRGESS algorithm successfully finds all the modes. Note also that the starting points of GMRGESS algorithms are drawn from Gaussian distribution with covariance matrix $40I_{2}$, which means EM-TMRGESS algorithm is more powerful to explore distant modes.


\begin{figure}[h]
\centering 
\includegraphics[height=3.5in]{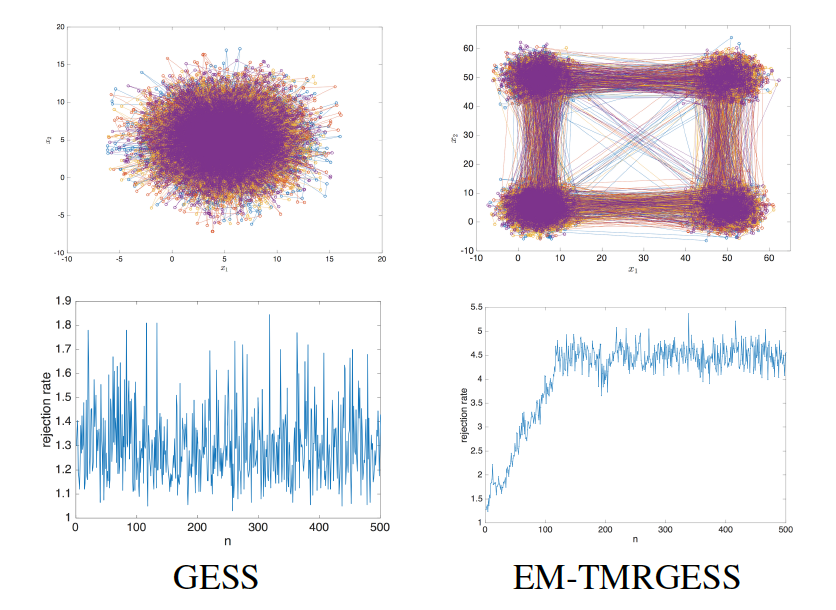}
\caption{From left to right: GESS algorithm and EM-TMRGESS algorithm. Plots of the first row show the drawn samples, in which two consecutive samples are connected with a line (0-125 iterations: blue; 125-250 iterations: brown; 250-375 iterations: orange; 375-500 iterations: purple). Plots of the second row show the rejection rates as a function of iteration number. } \label{frontier}
\end{figure}

\subsection{Forest CoverType classification}
In this subsection, we apply our algorithm to a forest cover type dataset $\mathnormal{covtype}$. The $\mathnormal{covtype}$ contains 495141 observations  with 55 variables, including one discrete variable as the response variable. Limited by the computing ability, we first filter out the data with response variable taking value of 0 or 1. Then we randomly select 4000 data from the filtered dataset and use the first 9 variables as independent variables. The data is standardized such that each variable has zero mean and unit variance. We then randomly split the data into 3000 training data and 1000 test data. Denote the training data as $\mathcal{D}_{1}=(X_{1}^{1},...,X_{3000}^{1},Y_{1}^{1},...,Y_{1000}^{1})$ and the test data as $\mathcal{D}_{2}=(X_{1}^{2},...,X_{3000}^{2},Y_{1}^{2},...,Y_{1000}^{2})$, where $Y_{i}^{j}$ is binary (0-1). Denote the parameters as $\mathbf{\beta}$. Our target is to sample from the logistic log-likelihood function:
\begin{equation}
\log L(\beta; \mathcal{D})= \sum_{n=1}^{2000} Y_{n}^{1} \log p_{n}+ (1-Y_{n}^{1})\log(1-p_{n}),
\end{equation}
where
\begin {equation}
p_{n}=\frac{1}{1+e^{-\mathbf{\beta} X_{n}^{1}}}
\end{equation}
In this example, we want to compare the prediction accuracy of EM-GMRGESS, VI-GMRGESS, SA-GMRGESS, EM-TMRGESS, GESS, MH sampling and Hamiltonian MCMC (HMC). We use `STAN' to conduct HMC sampling. For MH sampling, the proposal function is set to be Gaussian distribution centered at the previous state with covariance $10I_{9}$. The starting points are drawn from Gaussian distribution $\mathcal{N}(\cdot | 0, 5I_{9})$. We use 4-components mixture and 50 parallel chains. For our proposed algorithm and GESS, the total number of iterations is $10000$, $5000$ of which is set to be the burn-in period. For MH sampling and HMC, the number of iteration is 40000 and 1000 respectively. Denote $\hat{\beta}$ as the estimation for $\beta$ and $\widehat{Y_{n}^{2}}$ as the prediction for $Y_{n}^{2}$. Let $\hat{p_{n}}=\frac{1}{1+e^{-\hat{\beta} X_{n}^{2}}}$, then $\widehat{Y_{n}^{2}}= I\{\hat{p_{n}}>0.5\}$. The model accuracy for the test data is defined as:
\begin{equation}
Accuracy= \frac{ \sum_{n=1}^{2000} I\{ Y_{n}^{2}= \widehat{Y_{n}^{2}} \} } {2000}
\end{equation}

\begin{table}
    \centering
    \caption {Accuracy for the test data.}
    \begin{tabular}{ccc}
    \hline
    &  Accuracy & Time\\ \hline
    EM-GMRGESS & 0.585  &  0.039s  \\ \hline
    VI-GMRGESS  & 0.570  &  0.027s \\ \hline
    SA-GMRGESS  & 0.581  &  0.023s \\ \hline
    EM-TMRGESS  & 0.592 & 0.021s \\ \hline
    GESS  & 0.571 & 0.015s \\  \hline    
    MH sampling &     0.418    & 0.007s \\ \hline
    HMC  &   0.566  &  0.029s \\ \hline
    \end{tabular}
\end{table}

Table 1 shows results of different MCMC algorithms. The first column shows the prediction accuracy. According to the first column, the performance of our algorithms are better than HMC, GESS and MH algorithms. MH algorithm works worst because it can not converge. Explanation for the higher accuracy is that when the target distribution only has one mode, our algorithms has model averaging effect because the four-components mixture RGESS can be deemed as four GESS. The second column of Table 1 compares the average processing time to generate a new sample. The processing time of MH sampling is least of all but its performance is also the worst. GESS algorithm is faster than the proposed algorithms because it only tunes the parameters of one single Student's t-distribution. The processing time of proposed algorithms are similar to that of HMC, but the performance are relatively better. Figure 7 shows plots of samples as a function of iteration numbers on a single Markov chain. All algorithms except MH algorithm converge to some value. According to the results (table 1), we recommend using EM-TMRGESS in practice.

\begin{figure}[htb]
\centering 
\includegraphics[height=4in]{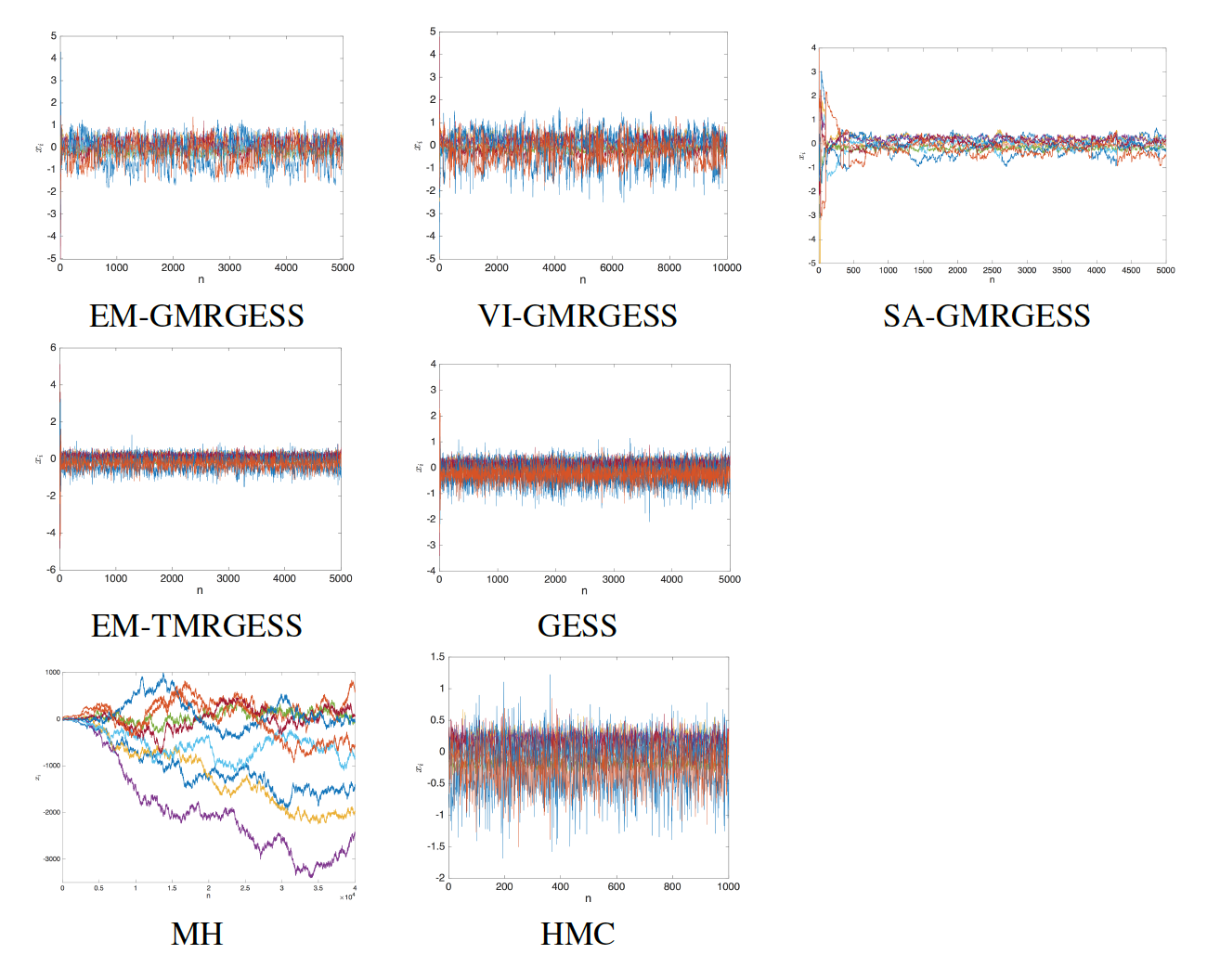}
\caption{Plots of the nine dimensions of samples on one single Markov chain, each color represents a specific dimension.  } \label{frontier}
\end{figure}

\begin{figure}[!h]
\centering 
\includegraphics[height=2in]{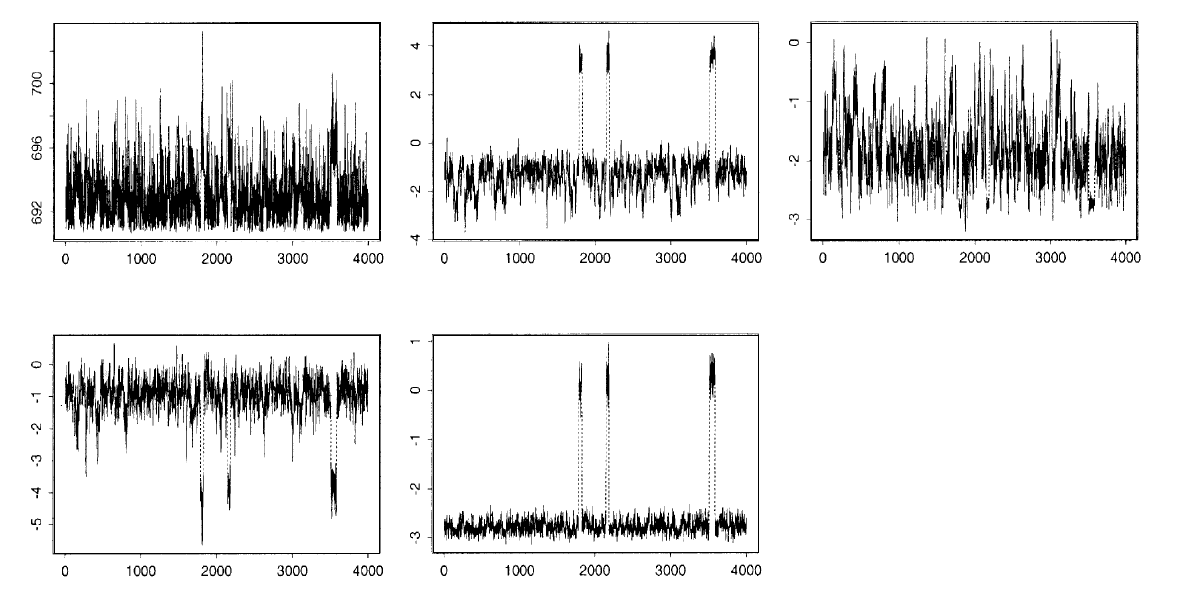}
\caption{Plots of energy (upper left), $\tilde{\gamma}$(upper middle), $\tilde{\mu}$ (upper right), $\tilde{\theta}$ (lower left), $\tilde{v}$ (lower middle) as a function of number of iterations. The plots are over a period of 4000 iterations (after convergence). Dashed lines indicate global moves.} \label{frontier}
\end{figure}


\begin{figure}[!h]
\centering 
\includegraphics[height=3.5in]{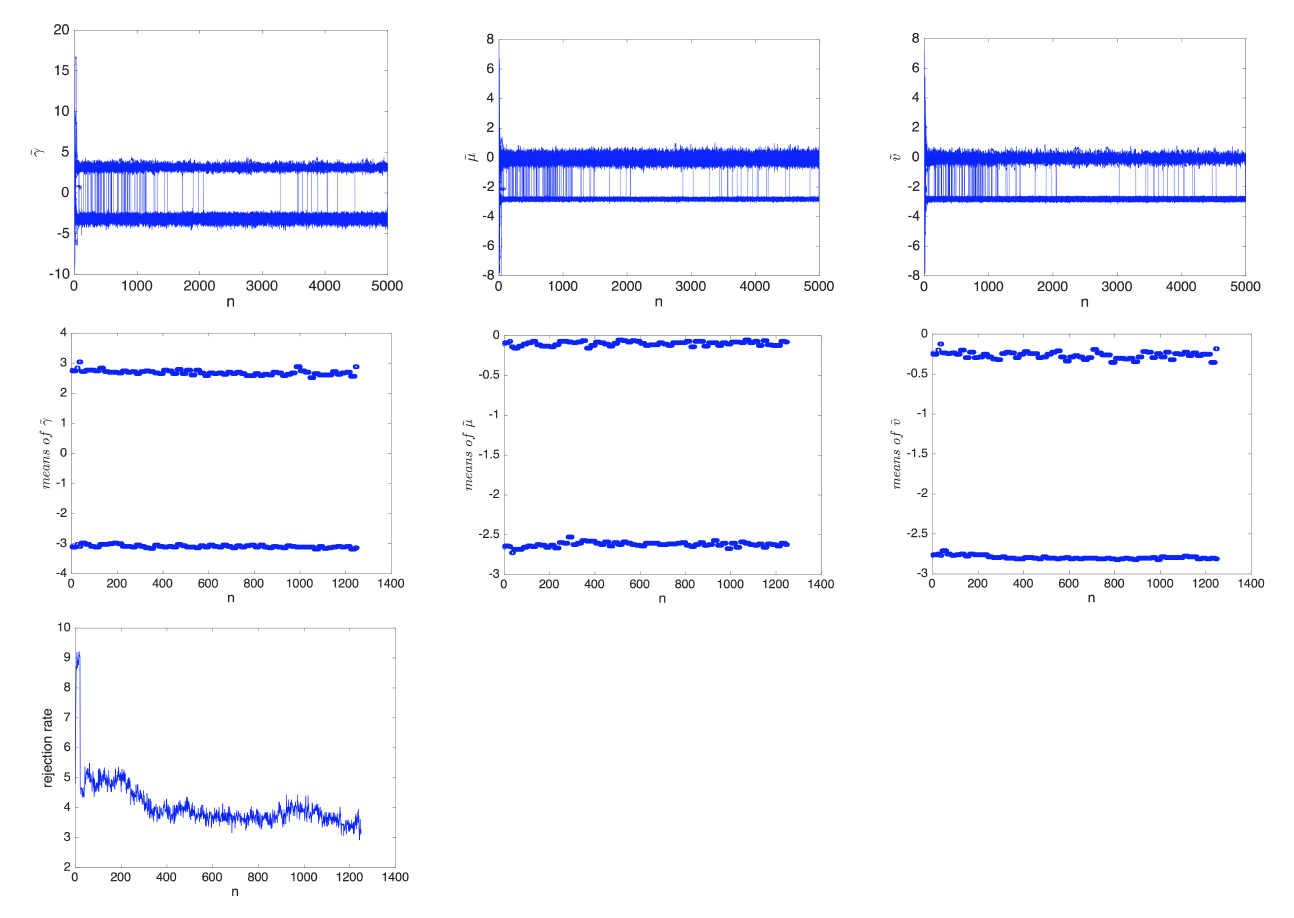}
\caption{EM-TMRGESS: plots of $\tilde{\gamma}$ (upper left), $\tilde{\mu}$(upper middle), $\tilde{v}$ (upper right), estimated means of $\tilde{\gamma}$ (middle left), estimated means of $\tilde{\mu}$ (middle middle), estimated means of $v$ (middle right), rejection rates (lower left). } \label{frontier}
\end{figure}

\begin{figure}[!h]
\centering 
\includegraphics[height=3.5in]{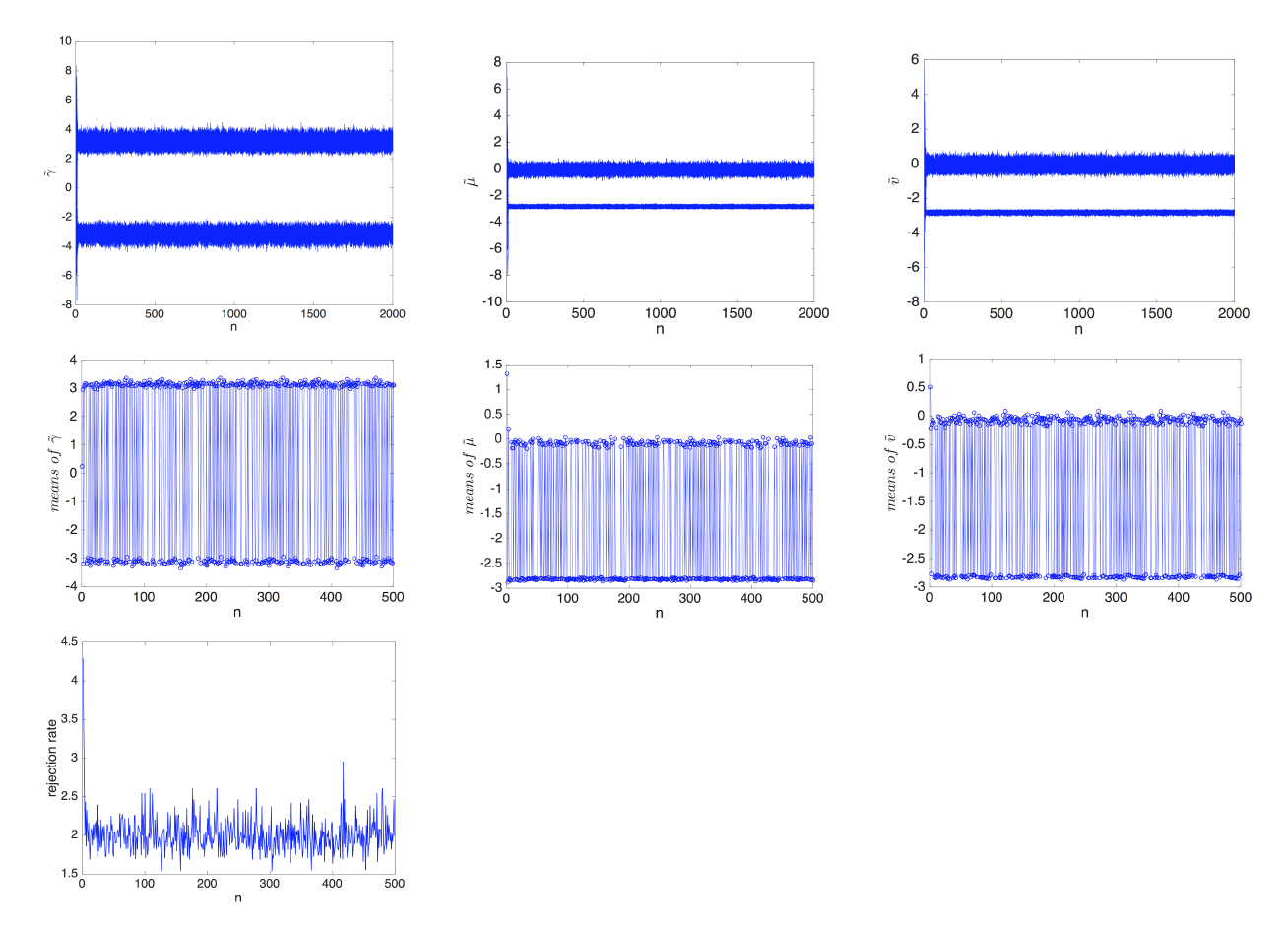}
\caption{EM-GMRGESS: plots of $\tilde{\gamma}$ (upper left), $\tilde{\mu}$ (upper middle), $\tilde{v}$ (upper right), estimated means of $\tilde{\gamma}$ (middle left), estimated means of $\tilde{\mu}$ (middle middle), estimated means of $v$ (middle right), rejection rates (lower left). } \label{frontier}
\end{figure}

\begin{figure}[!h]
\centering 
\includegraphics[height=3.5in]{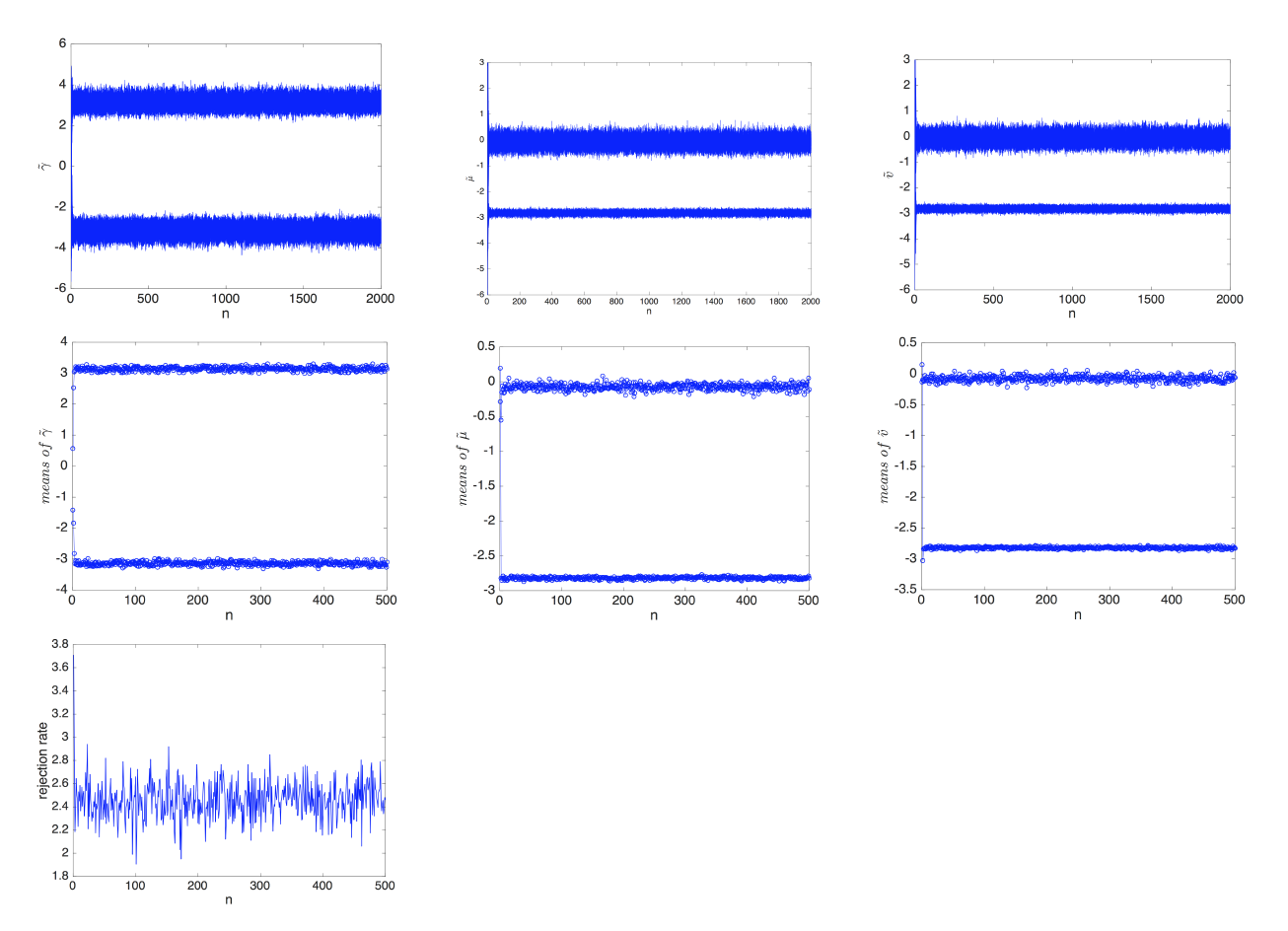}
\caption{VI-GMRGESS: plots of $\tilde{\gamma}$ (upper left), $\tilde{\mu}$ (upper middle), $\tilde{v}$ (upper right), estimated means of $\tilde{\gamma}$ (middle left), estimated means of $\tilde{\mu}$ (middle middle), estimated means of $v$ (middle right), rejection rates (lower left). } \label{frontier}
\end{figure}

\begin{figure}[!h]
\centering 
\includegraphics[height=3.5in]{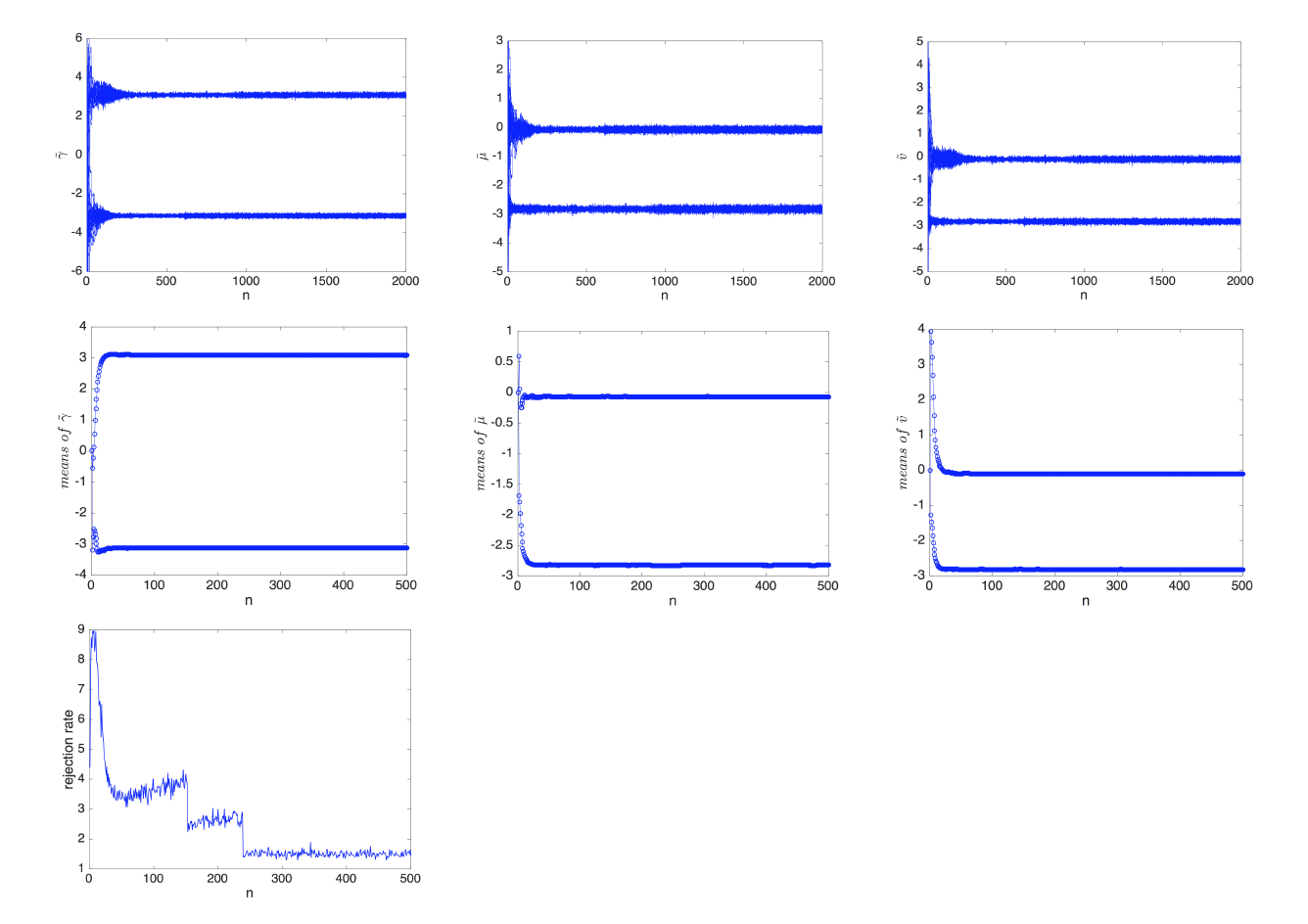}
\caption{SA-GMRGESS: plots of $\tilde{\gamma}$ (upper left), $\tilde{\mu}$ (upper middle), $\tilde{v}$ (upper right), estimated means of $\tilde{\gamma}$ (middle left), estimated means of $\tilde{\mu}$ (middle middle), estimated means of $v$ (middle right), rejection rates (lower left). } \label{frontier}
\end{figure}

\begin{table}
    \centering
    \caption {Number of dead fetuses in litters of mice. For each combination of litter size (n) and number of dead (x), the number of litters is given. No litters had more than nine dead.}
    \begin{tabular}{|c|c|c|c|c|c|c|c|c|c|c|}
    \hline
    & $\mathbf{0}$ & $\mathbf{1}$ & $\mathbf{2}$ & $\mathbf{3}$ & $\mathbf{4}$ &$\mathbf{5}$& $\mathbf{6}$&$\mathbf{7}$& $\mathbf{8}$ & $\mathbf{9}$ \\ \hline
    1&  7 & 0 &    &  & &  &  &  &  &   \\ \hline
    2&  7 & 0 & 0 &  & &  &  &  &  &   \\ \hline
    3&  6 & 0 & 0 & 0  & 0 &  &  &  &  &   \\ \hline
    4&  5 & 2 & 1 & 0  &0 &  &  &  &  &   \\ \hline
    5&   8&2  &1  &0  &1 & 1 &  &  &  &   \\ \hline
    6&   8&0  &0  & 0 &0 & 0 & 0 & 0 & 0 &   \\ \hline
    7&   4& 4 &2  &1  &0 & 0 & 0 & 0 &  &   \\ \hline
    8&   7& 7 & 1 & 0 & 0& 0 &0  & 0 & 0 &0   \\ \hline
    9&   8& 9 &7  &1  &1 & 0 & 0 & 0 & 0 & 0  \\ \hline
    10&  22 & 17 & 2 & 0 &1 & 0 & 0 & 1 & 1 & 0  \\ \hline
    11&   30& 18 & 9 & 1 & 2 & 0 & 1 & 0 & 1 & 0  \\ \hline
    12&   54& 27 & 12 & 2 &1 & 0 & 2 & 1 & 0 & 0  \\ \hline
    13&   46& 30 & 8 & 4 & 1 & 1 & 0 & 1 & 0 & 0  \\ \hline
    14&   43& 21 & 13 & 3 & 1& 0 & 0 & 1 & 0 &1   \\ \hline
    15&   22& 22 & 5 & 2 &1 & 0 & 0 & 0 & 0 & 0  \\ \hline
    16&   6& 6 & 3 &0  &1 & 1 & 0 & 0 & 0 & 0  \\ \hline
    17&   0&  0& 0 & 0 & 0& 0 & 0 & 0 & 0 & 0  \\ \hline
    18&   3& 0 & 2 & 1 &0 &0  & 0 & 0 & 0 & 0  \\ \hline
    \end{tabular}
\end{table}


\subsection{Fetal deaths in litters}
In this subsection, we apply our algorithms to a finite mixture model as used in \cite{Brooks1997Finite}. The finite mixture model is used to describe fetal deaths in litters of mice. In their paper, six datasets are presented and we choose the largest one (Table 3). Since the data are clearly over-dispersed, they used a mixture of beta-binomial model and a binomial distribution to fit the data. Two modes are estimated in their paper using maximum likelihood methods. \cite{Tjelmeland2001Mode} also studied this model by using an MCMC algorithm with optimization based proposals. Figure 8 is cited from the paper, their method can find the second mode indicated by the dashed lines, but with a very low frequency. In this example, we use a mixture of two binomial distributions as follows:
\begin{equation}
P(X=x|n) = \gamma \binom{n}{x} \mu^{x} (1-\mu)^{n-x} + (1-\gamma) \binom{n}{x} v^{x} (1-v)^{n-x}
\end{equation}
where the parameters $ \mu \in [0,1], v \in [0,1], \gamma \in [0,1] $. To enable the parameters to take values on all $\mathbb{R}$, we adopt logit transformations to $\mu, v$ and $ \gamma $, i.e.
\begin{equation}
\mu = \frac{exp(\tilde{\mu})}{1+exp(\tilde{\mu})}, ~~~ v = \frac{exp(\tilde{v})}{1+exp(\tilde{v})},~~~ \gamma = \frac{exp(\tilde{\gamma})}{1+exp(\tilde{\gamma})}
\end{equation}
where $\tilde{\mu}, \tilde{v}, \tilde{\gamma} \in \mathbb{R}^{1}$. 

To sample from equation (20),  we adapt the parameters every 20 iterations and draw 4 samples repeatedly at every iteration. The number of components of mixture distributions is set to be 2. The starting points for $\tilde{\mu}, \tilde{\gamma}$ and $\tilde{v}$ are drawn from Gaussian distribution with mean $0$ and covariance $5I_{3}$. Figure 9 are the plot of results as a function of iteration sampled from EM-TMRGESS algorithm. Figure 10 to Figure 12 are results of EM-GMRGESS, VI-GMRGESS and SA-GMRGESS algorithms. As shown in the plots, the proposed algorithms first find the two clearly separated modes, then adapt the parameters so as to reduce the rejection rates. For the EM-TMRGESS algorithm, samples can jump between the two modes because Student's t-distribution has long tails. Because of parameters adaption there are less and less interactions between two modes as the number of iteration $n$ increases, meanwhile the rejection rate is also decreasing.  For the EM-GMRGESS, VI-GMRGESS and SA-GMRGESS algorithms, samples do not jump between modes after successfully detecting the two modes. Because of the interaction between two modes, EM-TMRGESS algorithm has higher average rejection rate than the other algorithms. However, it is still much more efficient than the algorithm in \cite{Tjelmeland2001Mode}, whose average acceptance rate of mode jumping proposals is only $2.5 \%$ (equivalent to a rejection rate of 40). What's more, the estimated means of the parameters can give us additional information for locations of modes.

\section{Conclusion}
In this paper, we proposed several regional generalized elliptical slice sampling (RGESS) algorithms. For GMRGESS, we try EM algorithm and VI algorithm to adapt the parameters of the Gaussian mixture distribution, and we use stochastic approximation algorithm to minimize the KL-divergence between the Gaussian mixture distribution and the target distribution. For TMRGESS, we use EM algorithm to estimate the parameters of Student's t-mixture distribution. Theoretical proofs are given to show the ergodicity of above algorithms. The experimental results shows: with the same starting points our sampler can reach more distant modes; when doing Bayesian inference for a uni-modal posterior distribution, our algorithms can give more accurate estimations because of the model averaging effect; when doing Bayesian inference for a multi-modal posterior distribution, our algorithms can find the modes well and the estimated means of the mixture distribution can be used as additional information for the location of modes; our algorithms are more efficient with lower rejection rate.

There are already studies on adaptive MCMC and using mixture distribution as the proposal function, but to our knowledge we are the first to combine them with elliptical slice sampling and use regional priors. Still, there are some questions left for further research:

\begin{itemize}
\item We approximate the target distribution with Gaussian mixture distribution and Student's t-mixture distribution. This can be applied to many other studies. For example, in stochastic volatility (SV) models, usually we approximated the distribution of some variable by Gaussian mixture distribution so as to convert it to Gaussian state space model. 


\item Instead of using some criterion (such as AIC and BIC) to determine the number of mixture components, we just use a relatively large number for sampling accuracy. However, redundant components can impair sampling efficiency. Therefore, a better method is desired to determine the number of components so as to balance the sampling efficiency and accuracy.

\item We do not consider any strategy to determine when to stop the adaption. This sometimes can lead to so-called `over-adaption' problem (for example, there are four modes detected at $n_{th}$ iteration, but at following iterations there are only three modes detected). One possible situation for this to happen is when the samples used for adaption after $n_{th}$ iteration happen to gather around  3 modes, and the eigenvalues of the estimated covariance is small. 

\end{itemize}

\section*{References}
\bibliography{ref.bib}
\newpage

\begin{appendices}

\section{Proof of detailed balance}


\subsection{Proof of Theorem 1}
The proposal function in equation (5) is $f(x_{2}|x_{1})=\sum_{i=1}^{M}\mathcal{I}(x_{1}\in S_{i})f_{i}(x_{2})$. \\
If $x_{1} \in S_{i}, x_{2} \in S_{j}$, then $f(x_{2}|x_{1}) = f_{i}(x_{2}), f(x_{1}|x_{2}) = f_{j}(x_{1})$. The acceptance rate is:
\begin{equation}
\frac{\pi(x_{2})f(x_{1}|x_{2})}{\pi(x_{1})f_{i}(x_{2}|x_{1})} = \frac{\pi(x_{2})f_{j}(x_{1})}{\pi(x_{1})f_{i}(x_{2})}
\end{equation}
If $x_{1} \in S_{i}, x_{2} \in S_{i}$, then $f(x_{2}|x_{1}) = f_{i}(x_{2}), f(x_{1}|x_{2}) = f_{i}(x_{1})$, the acceptance rate is
\begin{equation}
\frac{\pi(x_{2})f(x_{1}|x_{2})}{\pi(x_{1})f_{i}(x_{2}|x_{1})} = \frac{\pi(x_{2})}{\pi(x_{1})}
\end{equation}

\subsection{Proof of Theorem 2}
As shown in Section 3.1, ESS and GESS algorithms are essentially the same as MH algorithm, the pseudo-prior of GESS plays the same role as the proposal function in MH. Therefore, we deem the transition proposal $f_{j}(x_{1}|x_{2})$ (the probability of transiting from $x_{2}$ to $x_{1}$ when $x_{2} \in S_{j}$) equals the pseudo-prior $\mathcal{N}(x_{2}; \mu_{j}, \Sigma_{j})$ when $x_{2} \in S_{j}$.  In this way, one only needs to prove $\frac{\pi(x_{2})f_{j}(x_{1}|x_{2})}{\pi(x_{1})f_{i}(x_{2}|x_{1})} = \frac{R_{i}(x_{2})}{R_{j}(x_{1})}$, if  $x_{2} \in S_{j}, x_{1} \in S_{i}$. 
\begin{equation}
\frac{\pi(x_{2})f_{j}(x_{1}|x_{2})}{\pi(x_{1})f_{i}(x_{2}|x_{1})} = \frac{\pi(x_{2})\mathcal{N}(x_{1}; \mu_{j}, \Sigma_{j})}{\pi(x_{1})\mathcal{N}(x_{2}; \mu_{i}, \Sigma_{i})} = \frac{R_{i}(x_{2})}{R_{j}(x_{1})}
\end{equation}

\subsection{Proof of Theorem 3}
As shown in Section 3.1, ESS and GESS algorithms are essentially the same as MH algorithm, the pseudo-prior of GESS plays the same role as the proposal function in MH. Therefore, we deem the transition proposal $f_{j}(x_{1}|x_{2})$ (the probability of transiting from $x_{2}$ to $x_{1}$ when $x_{2} \in S_{j}$) equals the pseudo-prior $\mathcal{T}(x_{2}; \mu_{j}, \Sigma_{j}, \nu_{j})$ when $x_{2} \in S_{j}$.  In this way, one only needs to prove $\frac{\pi(x_{2})f_{j}(x_{1}|x_{2})}{\pi(x_{1})f_{i}(x_{2}|x_{1})} = \frac{R_{i}(x_{2})}{R_{j}(x_{1})}$, if  $x_{2} \in S_{j}, x_{1} \in S_{i}$. 
\begin{equation}
\frac{\pi(x_{2})f_{j}(x_{1}|x_{2})}{\pi(x_{1})f_{i}(x_{2}|x_{1})} = \frac{\pi(x_{2})\mathcal{T}(x_{1}; \mu_{j}, \Sigma_{j}, \nu_{j})}{\pi(x_{1})\mathcal{T}(x_{2}; \mu_{i}, \Sigma_{i}, \nu_{i})} = \frac{R_{i}(x_{2})}{R_{j}(x_{1})}
\end{equation}


\section{Stochastic approximation algorithm for Gaussian mixture distribution estimation}
Here we show the stochastic approximation algorithm to minimize the KL-divergence between the Gaussian mixture distribution and the target distribution. Denote the target distribution as $\pi(\mathbf{x})$ and the Gaussian mixture distribution as $f(\mathbf{x}; \phi)$, where $f(\mathbf{x}; \phi)=\sum_{m=1}^{M}w_{m}\mathcal{N}(\mathbf{x}|\mathbf{\mu}_{m},\Sigma_{m})$. We wish to find $\phi^{\ast}$ that minimizes the KL divergence $\mathcal{D}[\pi(\mathbf{x}) \lVert f(\mathbf{x},\phi)]=\mathbb{E}_{\pi}[log \frac{\pi(\mathbf{x})}{f(\mathbf{x},\phi)}]$. Therefore, $\phi^{\ast}$ is the root of 

\begin{equation}
g(\phi)=\int \frac{\pi(\mathbf{x})}{f(\mathbf{x};\phi)}\frac{\partial}{\partial \phi} f(\mathbf{x},\phi)=0.
\end{equation}

We apply the Stochastic Approximation (SA) to solve above equation iteratively. Denote $G(\mathbf{\mathbf{x}},\phi)=\frac{\partial}{\partial \phi}[\log \frac{\pi(\mathbf{x})}{f(\mathbf{x},\phi)}]$, then $g(X^{(1:K)},\phi) \approx \frac{1}{K}\sum_{k=1}^{K}G(X^{k},\phi)$ where $X^{k} \sim \pi(\mathbf{x})$. Let $\hat{g}(X^{(1:K)};\phi)=\frac{1}{K}\sum_{k=1}^{K}G(X^{k},\phi)$, then $\hat{g}(X^{(1:K)};\phi)$ is the estimate of $g(X^{(1:K)},\phi)$, such that\\ $E(\hat{g}(X^{(1:K)};\phi))=g(X^{(1:K)},\phi)$. The stochastic approximation iteration is
\begin{equation}
\begin{split}
\phi_{n+1}&=\phi_{n}-r_{n+1}(0-(\hat{g}(X^{(1:K)};\phi)))\\&=\phi_{n}+r_{n+1}\hat{g}(X^{(1:K)};\phi)
\end{split}
\end{equation}
In our case, the SA algorithm is equivalent to a gradient descent algorithm, with the gradient at each iteration approximated by the Monte Carlo method. The update equations can be easily calculated as:
\begin{equation}
\begin{split}
w_{j}^{n+1}=&w_{j}^{n}+r_{n+1}[\frac{1}{N}\sum_{k=1}^{K}\frac{\mathcal{N}(X^{n}_{k}|\mu_{j}^{n},\Sigma_{j}^{n})}{\sum_{m=1}^{M}w_{m}\mathcal{N}(X^{n}_{k}|\mathbf{\mu}_{m}^{n},\Sigma_{m}^{n})}\\
&-\frac{1}{MK}\sum_{k=1}^{K}\sum_{m=1}^{M}\frac{\mathcal{N}(X^{k}_{n}|\mu_{m}^{n},\Sigma_{m}^{n})}{\sum_{i=1}^{M}w_{i}^{n}\mathcal{N}(X^{n}_{k}|\mathbf{\mu}_{i}^{n},\Sigma_{i}^{n})}]
\end{split}
\end{equation}
\begin{equation}
\begin{split}
\mu_{j}^{n+1}=&\mathbf{\mu}_{j}^{n}+r_{n+1}\frac{1}{K}\sum_{k=1}^{K}\frac{\mathcal{N}(X^{n}_{k}|\mu_{j}^{n},\Sigma_{j}^{n})}{\sum_{m=1}^{M}w_{m}^{n}\mathcal{N}(X^{n}_{k}|\mathbf{\mu}_{m}^{n},\Sigma_{m}^{n})}\times \\& (\Sigma_{m}^{n})^{-1}(X^{n}_{k}-\mu_{m}^{n})
\end{split}
\end{equation}
\begin{equation}
\begin{split}
\Sigma_{j}^{n+1}=&\Sigma_{j}^{n}+r_{n+1}\frac{1}{K}\sum_{k=1}^{K}\frac{\mathcal{N}(X^{n}_{k}|\mu_{j}^{n},\Sigma_{j}^{n})}{\sum_{m=1}^{M}w_{m}^{n}\mathcal{N}(X^{n}_{k}|\mathbf{\mu}_{m}^{n},\Sigma_{m}^{n})}\times \\&[(X^{n}_{k}-\mu_{m}^{n})(X_{k}^{n}-\mu_{m}^{n})^{T}-\Sigma_{m}^{n}]
\end{split}
\end{equation}

\section{Proofs of theorems in section 6}
\subsection{Ergodicity of EM-RGESS}
Proof: 
Our proof is mainly based on \cite{craiu2009learn} because GESS is essentially a kind of MH algorithm that has adaptive step size as well as the residual function as the proposal function. In this way, we need to prove that the proposed algorithms satisfy the simultaneous uniform ergodicity and diminishing adaption conditions.  

(a) [Simultaneous uniform ergodicity]

Since S is compact, by positivity and continuity, let $d=sup_{x\in\mathcal{S}}\pi(x)<+\infty$ and $\epsilon=\min_{x\in\mathcal{S},m\leq M}\mathcal{N}(x|\mu_{m},\Sigma_{m})>0$. The regional proposal function can be written as:
\begin{equation}
q_{\gamma}(x,y) = \sum_{m=1}^{M}\mathbb{I}_{\mathcal{S}_{m}}(x)\mathcal{N}(y|\mu_{m},\Sigma_{m}) \geq \epsilon
\end{equation}
For $x\in \mathcal{S}$ and $B \subset \mathcal{S}$, let
\begin{equation}
R_{x,\gamma}(B) = \{y \in B: \frac{\pi(y)q_{\gamma}(y,x)}{\pi(x)q_{\gamma}(x,y)}<1\}
\end{equation} 
and $A_{x,\gamma}(B) = B \setminus R_{x,\gamma}(B)$. Then we have
\begin{equation}
\begin{split}
P_{\gamma}(x,B) \geq& \int_{R_{x,\gamma}(B)} q_{\gamma}(x,y) \min \{ \frac{\pi(y)q_{\gamma}(y,x)}{\pi(x)q_{\gamma}(x,y)},1\} \mu^{Leb} dy\\ + & \int_{A_{x,\gamma}(B)} q_{\gamma}(x,y) \min \{ \frac{\pi(y)q_{\gamma}(y,x)}{\pi(x)q_{\gamma}(x,y)},1\} \mu^{Leb} dy \\= & \int_{R_{x,\gamma}(B)} \frac{\pi(y)q_{\gamma}(y,x)}{\pi(x)} \mu^{Leb} dy + \int_{A_{x,\gamma}(B)} q_{\gamma}(y,x) \mu^{Leb} dy \\ \geq & \frac{\epsilon}{d} \int_{R_{x,\gamma}(B)} \pi(y)\mu^{Leb} dy + \frac{\epsilon}{d} \int_{A_{x,\gamma}(B)} \pi(y)\mu^{Leb} dy = \frac{\epsilon}{d} \pi(B)
\end{split}
\end{equation}
Thus $P_{\gamma}(x,B) \geq v(B)$, where $v(B)=\frac{\epsilon}{d}\pi(B)$ is a nontrivial measure on $\mathcal{S}$. Then it follows from theorem 16.02 of \cite{meyn2012markov} that there is a $\rho=1-v(\mathcal{S})=1-\frac{\epsilon}{d}$ s.t. $|P_{\gamma}^{n}(x,A) - \pi(A)| \leq \rho^{n} $. Thus the simultaneous uniform ergodicity holds.

(b) [Diminishing adaption]\\
It is straightforward to prove that for any $ B \subset \mathcal{B}(\mathcal{S}), \lim_{n \to \infty} \sup_{x \in \mathcal{S}} || P_{\Gamma_{n+1}}(x,B) - P_{\Gamma_{n}}(x,B) || = 0$ because of the convergence of EM algorithm as $n$ goes to infinity. Next we prove it rigorously. Suppose $x \in \mathcal{S}_{1}, A \in \mathcal{B}(\mathcal{S})$. The proposal function is $f_{\lambda^{k}}(x,y)=\sum_{i=1}^{M}\mathbb{I}_{S_{i}}(x)\mathcal{N}(y|\mu_{i}^{k},\Sigma_{i}^{k})$.
\begin{equation}
\begin{split}
P_{\gamma_{k}}(x,A) =& \sum_{i \neq 1}\int_{A\cap S_{i}}{\mathcal{N}(y|\mu_{1}^{k},\Sigma_{1}^{k})} \min \{ 1, \frac{\pi(y)\mathcal{N}(x|\mu_{i}^{k},\Sigma_{i}^{k})}{\pi(x)\mathcal{N}(y|\mu_{1}^{k},\Sigma_{1}^{k})} \}dy \\+&\int_{A\cap S_{1}}{\mathcal{N}(y|\mu_{1}^{k},\Sigma_{1}^{k})} \min \{ 1, \frac{\pi(y)}{\pi(x)} \}dy \\+ & \delta_{x}(A) \int_{S_{1}}{\mathcal{N}(y|\mu_{1}^{k},\Sigma_{1}^{k})} (1- \min \{ 1, \frac{\pi(y)}{\pi(x)} \})dy \\+&  \delta_{x}(A) \sum_{i \neq 1}\int_{A\cap S_{i}}{\mathcal{N}(y|\mu_{1}^{k},\Sigma_{1}^{k})} \min (1-\{ 1, \frac{\pi(y)\mathcal{N}(y|\mu_{i}^{k},\Sigma_{i}^{k})}{\pi(x)\mathcal{N}(y|\mu_{1}^{k},\Sigma_{1}^{k})} \})dy 
\end{split}
\end{equation}
where $\delta_{x}(A)$ equals 1 if $x \in A$ and 0 otherwise. Denote the first term as $\uppercase\expandafter{\romannumeral1}_{\gamma_{k}}(x,A)$, second term as $\uppercase\expandafter{\romannumeral2}_{\gamma_{k}}(x,A)$, third term as $\uppercase\expandafter{\romannumeral3}_{\gamma_{k}}(x,A)$, forth term as $\uppercase\expandafter{\romannumeral4}_{\gamma_{k}}(x,A)$. 
\begin{equation}
\begin{split}
|P_{\gamma_{k+1}}(x,A) - P_{\gamma_{k}}(x,A)| \leq& |\uppercase\expandafter{\romannumeral1}_{\gamma_{k+1}}(x,A)-\uppercase\expandafter{\romannumeral1}_{\gamma_{k}}(x,A)| + |\uppercase\expandafter{\romannumeral2}_{\gamma_{k+1}}(x,A)-\uppercase\expandafter{\romannumeral2}_{\gamma_{k}}(x,A)|| \\+&\uppercase\expandafter{\romannumeral3}_{\gamma_{k+1}}(x,A)-\uppercase\expandafter{\romannumeral3}_{\gamma_{k}}(x,A)|+|\uppercase\expandafter{\romannumeral4}_{\gamma_{k+1}}(x,A)-\uppercase\expandafter{\romannumeral4}_{\gamma_{k}}(x,A)|
\end{split}
\end{equation}
Next we prove that $\lim_{k \to \infty}|\uppercase\expandafter{\romannumeral1}_{\gamma_{k+1}}(x,A)-\uppercase\expandafter{\romannumeral1}_{\gamma_{k}}(x,A)|=0$. \\
Let $\alpha_{k}^{i}(x,y) = \min \{ 1, \frac{\pi(y)\mathcal{N}(y|\mu_{i}^{k},\Sigma_{i}^{k})}{\pi(x)\mathcal{N}(y|\mu_{1}^{k},\Sigma_{1}^{k})} \}$, then
\begin{equation}
\begin{split}
|\uppercase\expandafter{\romannumeral1}_{\gamma_{k+1}}(x,A)-\uppercase\expandafter{\romannumeral1}_{\gamma_{k}}(x,A)| =& \sum_{i \neq 1} \int_{A \cap S_{i}}| \mathcal{N}(y|\mu_{1}^{k+1},\Sigma_{1}^{k+1}) \alpha_{k+1}^{i}(x,y) - \mathcal{N}(y|\mu_{1}^{k},\Sigma_{1}^{k}) \alpha_{k}^{i}(x,y) |dy \\ \leq& \sum_{i \neq 1} \int_{A \cap S_{i}}| \mathcal{N}(y|\mu_{1}^{k+1},\Sigma_{1}^{k+1}) \alpha_{k+1}^{i}(x,y) - \mathcal{N}(y|\mu_{1}^{k+1},\Sigma_{1}^{k+1}) \alpha_{k}^{i}(x,y) |dy \\ +& \sum_{i \neq 1} \int_{A \cap S_{i}}| \mathcal{N}(y|\mu_{1}^{k+1},\Sigma_{1}^{k+1}) \alpha_{k}^{i}(x,y) - \mathcal{N}(y|\mu_ {1}^{k},\Sigma_{1}^{k}) \alpha_{k}^{i}(x,y) |dy \\ \leq & M \sum_{i \neq 1} \int_{A \cap S_{i}} |\alpha_{k+1}^{i}(x,y) - \alpha_{k}^{i}(x,y) |dy \\+& \sum_{i \neq 1} \int_{A \cap S_{i}} |\mathcal{N}(y|\mu_{1}^{k+1},\Sigma_{1}^{k+1})-\mathcal{N}(y|\mu_{1}^{k}, \Sigma_{1}^{k})|
\end{split}
\end{equation}
\begin{equation}
\begin{split}
M \sum_{i \neq 1} \int_{A \cap S_{i}} |\alpha_{k+1}^{i}(x,y) - \alpha_{k}^{i}(x,y) |dy =& M \sum_{i \neq 1} \int_{A \cap S_{i}}| \frac{\pi(y)\mathcal{N}(y|\mu_{i}^{k+1},\Sigma_{i}^{k+1})}{\pi(x)\mathcal{N}(y|\mu_{1}^{k+1},\Sigma_{1}^{k+1})} - \frac{\pi(y)\mathcal{N}(y|\mu_{i}^{k},\Sigma_{i}^{k})}{\pi(x)\mathcal{N}(y|\mu_{1}^{k},\Sigma_{1}^{k})} | dy \\& \leq \frac{Md}{\pi(x)} \sum_{i \neq 1} \int_{A \cap S_{i}} | \frac{\mathcal{N}(y|\mu_{i}^{k+1}, \Sigma_{i}^{k+1})}{\mathcal{N}(y|\mu_{1}^{k+1}, \Sigma_{1}^{k+1})} -\frac{\mathcal{N}(y|\mu_{i}^{k}, \Sigma_{i}^{k})}{\mathcal{N}(y|\mu_{1}^{k}, \Sigma_{1}^{k})} | dy
\end{split}
\end{equation}
where $M=\max_{y \in \mathcal{S},\mu,\Sigma} \mathcal{N}(y|\mu,\Sigma)$(assume $\mu, \Sigma$ are bounded), $d=\sup_{x \in \mathcal{S}} \pi(x) < \infty$.
Now we discuss the convergence of EM algorithm. Convergence of EM algorithm is guaranteed for a fixed dataset. For the proposed algorithm, the samples used for EM algorithm at different iterations are different. Since the samples follow the true target distribution after some iterations, here we make a reasonable but somewhat strong assumption that the estimates of EM algorithm on the samples converge after some iterations. $ \mu_{i}^{k+1} \to \mu_{i}^{k},\Sigma_{i}^{k+1} \to \Sigma_{i}^{k}$ as $k \to \infty$. Therefore, $|\uppercase\expandafter{\romannumeral1}_{\gamma_{k+1}}(x,A)-\uppercase\expandafter{\romannumeral1}_{\gamma_{k}}(x,A)| \to 0 $ as $ k \to \infty$. \\
Similarly, $|\uppercase\expandafter{\romannumeral2}_{\gamma_{k+1}}(x,A)-\uppercase\expandafter{\romannumeral2}_{\gamma_{k}}(x,A)| \to 0$, $|\uppercase\expandafter{\romannumeral3}_{\gamma_{k+1}}(x,A)-\uppercase\expandafter{\romannumeral3}_{\gamma_{k}}(x,A)| \to 0 $,$|\uppercase\expandafter{\romannumeral4}_{\gamma_{k+1}}(x,A)-\uppercase\expandafter{\romannumeral4}_{\gamma_{k}}(x,A)| \to 0 $. 

\subsection{Ergodicity of VI-RESS, SA-RESS, TM-RGESS}
We can make assumption that  the variational inference (VI), stochastic approximation (SA) for Gaussian mixture distribution and EM algorithm for Student's t-mixture distribution all converge after some iterations. The proofs are the same as the proof in Appendix C.1.

\end{appendices}
\end{document}